\documentclass{article}
\usepackage{graphicx} 
\usepackage[right]{lineno}
\usepackage[colorlinks,bookmarksopen,bookmarksnumbered,citecolor=red,urlcolor=red]{hyperref}
\usepackage{graphicx}
\usepackage[sort,numbers]{natbib}
\usepackage{lmodern}
\usepackage{amsmath}
\usepackage{epstopdf}
\usepackage{subfigure}
\usepackage{bbold}
\usepackage{booktabs}
\usepackage{etoolbox}
\usepackage{placeins}
\usepackage{fontawesome5}
\usepackage{caption} 
\usepackage{authblk}

\title{Hybrid Reduced-Order Models for Turbulent Flows Using Recurrent Neural Architectures}
\author{Haroon Imtiaz$^{1}$, Imran Akhtar$^{1}$, Muhammad R. Hajj$^{2}$}
\date{
    $^{1}$Department of Mechanical Engineering, NUST College of Electrical \& Mechanical Engineering, National University of Sciences \& Technology, Islamabad, Pakistan 44000. \\
    $^{2}$Department of Civil, Environmental and Ocean Engineering, Davidson Laboratory, Stevens Institute of Technology, 711 Hudson St, Hoboken, NJ 07030, USA. \\
}

\begin{document}
\maketitle

\begin{abstract}
{Proper-orthogonal decomposition (POD) based reduced-order models (ROM) of structurally dominant fluid flow can support a wide range of engineering applications. Yet, although they perform well for unsteady laminar flows, their straightforward extension to turbulent flows fails to capture the effects of small scale eddies and often leads to divergent solutions. Several approaches to mimic nonlinear closure terms modeling techniques within ROM frameworks have been employed to include the effect of higher modes that are often neglected. Recent success of neural network based models show promising results in modeling the effects of turbulence.
In this study, we augment POD-ROM with a recurrent neural network (RNN) to develop ROM for turbulent flows. We simulate a three dimensional flow past a circular cylinder at Reynolds number of 1000. We first compute the POD modes and project the Navier-Stokes equations onto the limited number of modes in a Galerkin approach to develop a conventional ROM and LES-inspired ROM for comparison.
We then develop a hybrid model by integrating the output of Galerkin projection ROM and long short-term memory (LSTM) RNN and term it as a physics-guided machine learning (PGML) model. The novelty of this study is to introduce a hybrid model that integrates LES inspired ROM and RNN to achieve more accurate and reliable predictions of turbulent flows. The results demonstrate that PGML for higher temporal coefficients outperforms the conventional and LES-inspired ROM.
}
\end{abstract}

\section{Introduction}
Reduced-order models (ROM) are often employed in fluid mechanics for design, optimization, and control purposes. Proper-orthogonal decomposition (POD) based ROM are usually applied in structurally dominant fluid flows.
Particularly, POD modes via dimensionality reduction method provide reduced bases for compact description of the flow \cite[]{taira2017modal}, whereas ROM describes the spatiotemporal evolution of a dynamical system.

ROM developed by Galerkin projection of the governing equations onto the POD modes works quite well for unsteady laminar flows. However, its application to turbulent flows poses multiple challenges due to the presence of higher modes in the flow field \cite{akhtar2010large,imtiaz2014closure,imtiaz2015closure,imtiaz2015proper}. This led to the modeling of a closure term within ROM framework inspired by large-eddy simulation (LES) technique in turbulence modeling \cite[]{imtiaz2020nonlinear, SROM, akhtar2012new, aubry1988dynamics, wang2011two, Borg, akhtar2010closure, maulik2019sub,maulik2019subgrid}. For instance,  Aubry et al. \cite{aubry1988dynamics} introduced the first closure term in ROM by using Heisenberg's model \cite{heisenberg1948statistischen} and investigated the wall region of a turbulent boundary layer.  Later, Couplet et al. \cite{Couplet2003} validated the use of LES-inspired closure models in ROM for complex flows because they noticed the energy transfer from low-index POD modes to high-index POD modes. Rempfer et al. \cite{rempfer1994evolution} improved the linear closure model by using the mode-dependent eddy viscosity approximation. Similarly, Borggaard et al. \cite{SROM} proposed two nonlinear closure terms in ROM inspired from LES. These models are termed as dynamic subgrid-scale model and variational multiscale model. Wang et al. \cite{wang2012proper} tested these models for turbulent flows and compared their performance with direct numerical simulation (DNS). Although the performance of ROM improved in terms of accuracy, however, its computational cost was relatively higher due to the presence of strain rate tensor. Akhtar et al. \cite{akhtar2012new} proposed a novel closure strategy in order to overcome the computational challenge of nonlinear closure term in ROM. Although this model is based on Smagorinsky closure model, it involves Frobenius matrix norm of the Jacobian of the conventional ROM. Thus, the closure model can be computed at a negligible cost for each time step. They tested this closure model for Burgers' equation and demonstrated its better performance in terms of  both accuracy and computational cost \cite{imtiaz2017closure}. Imtiaz and Akhtar \cite{imtiaz2020nonlinear} extended this approach to the Navier-Stokes equations to model the flow past a circular cylinder. It is important to note that the ROM of turbulent flow employed only the first few POD modes \cite{wang2012proper, imtiaz2020nonlinear}. Although, higher POD modes contain less amount of {\it{energy}} yet they play a vital role in accurately predicting the flow physics of the problem. Therefore, an accurate prediction of the temporal coefficients in the ROM is essential.

Recent advancements in data science, particularly in machine learning, have renewed interest in extracting more information from existing turbulence data. Machine learning can be categorized into three categories: supervised, semi-supervised, and unsupervised learning algorithms. Supervised and semi-supervised algorithms involve learning from expert and partial label data. Neural networks \cite[]{hornik1989multilayer, mjalled2023reduced} and regression methods \cite[]{zhang2015machine} are the examples of supervised learning, whereas reinforcement learning is a technique in semi-supervised learning \cite[]{sutton1999reinforcement}. In fluid mechanics, the neural networks can be employed to model convective heat transfer \cite[]{jambunathan1996evaluating}, fluid flows \cite[]{milano2002neural, hocevar2004experimental, maulik2020probabilistic}, and turbo-machinery \cite[]{pierret1999turbomachinery}. Since neural networks are nonlinear approximators, multiple layers in deep neural networks can be used for a wide range of functions and systems \cite[]{bcskei2019optimal,pawar2021data}.

POD is an unsupervised dimension reduction tool in fluid mechanics \cite[]{lumey2012stochastic} that is usually implemented to extract orthogonal and energy optimal modes from the flow field data \cite{noack1994low,berkooz1993proper,akhtar2009stability,rahman2018hybrid}. Sirovich \cite[]{Sirovich1987A, sirovich1987low} introduced the method of snapshots for computing POD modes via a simple data-driven procedure involving singular value decomposition. Since POD modes represent coherent structures in the flow field, they can be used for developing ROM of unsteady flows \cite[]{imtiaz2017lift, Akhtar2010A, akhtar2015using, Borggard_2016, fasel2022flexwing, callaham2022role} and efficient control design algorithm via neural networks \cite[]{gillies1998low, noack2013closed, duriez2014closed, brunton2015closed, vinuesa2024perspectives}.

Apart from the conventional ROM techniques, machine learning techniques are also employed for the development of ROM \cite{solera2024beta,san2018extreme}. For instance, recurrent neural network (RNN) can be used for time series forecasting in fluid mechanics. However, the prediction capability of RNNs is limited due to diminishing or exploding gradients that emerge during the training process. The development of long short-term memory (LSTM) has renewed the interest in RNN \cite[]{hochreiter1997long,pawar2020data, ahmed2020long,ahmed2021nudged,pawar2020long,kherad2024surrogate, saeed2023deep, zahn2024prediction} because it employs cell states and the gating mechanism for storing and forgetting the information during the training process. Thus, it can be used for predicting the long-term nonlinear behavior of turbulent flows. Rehman et al. \cite{rahman2019nonintrusive} presented nonintrusive ROM by using LSTM and employed it for predicting the temporal coefficients of the 2D quasi-geographic turbulence problem. This pure data-driven model provides reasonable results for long-time prediction of temporal coefficients, however it requires extensive training data because the model's generalizability is a great challenge.

Pawar et al. \cite{pawar2021model} introduced a physics-guided machine learning (PGML) framework where LSTM and POD-ROM are fused to improve the generalizability of ROM. However, the study was limited to 2D cases. In this study, we test the proposed framework for the 3D flow past a circular cylinder at Re = 1000. We compare critical parameters including temporal coefficients, velocity fields, Reynolds stresses, and kinetic energy spectra. Particularly, we compare the accuracy of the PGML model with conventional and LES-inspired ROM. The key contributions are to introduce a hybrid model of LES-inspired ROM and RNN to achieve more accurate and reliable predictions of turbulent flows.

The manuscript is organized as follows: Section 2 introduces the numerical methodology and presents the velocity field for the 3D flow past a cylinder. It demonstrates that the wake is characterized by turbulence at Reynolds number of 1000. Section 3 discusses the procedure for computing the POD modes and presents the modal structure of the velocity fields in the wake. In Section 4, we develop the ROM by using Galerkin projection methods and highlight its limitations in the context of turbulent flows. We introduce the hybrid framework for PGML model in detail in Section 5. Section 6 compares the performance of the PGML model with the conventional and LES-inspired models. Section 7 concludes the findings of this work.

\section{Numerical Methodology}

The governing equations for the incompressible flow field are the continuity equation and Navier-Stokes equations, which are written in a non-dimensional form as follows:
\begin{equation}
\left. \begin{array}{l}
\nabla \cdot \mathbf{u} = 0 \\
{\mathbf{u}_t} - {{\mathop{\rm Re}\nolimits} ^{ - 1}}\Delta \mathbf{u} + \left( {\mathbf{u}\cdot\nabla } \right)\mathbf{u} + \nabla p = 0
\end{array} \right\}
\label{eqn:NSE}
\end{equation}
where $\mathbf{u}$ and $p$ are the velocity and pressure fields, respectively. Here, the Reynolds number is defined as $\rm{Re}$ = $U_{\infty}$D/$\nu_D$ where D is the diameter of the circular cylinder, $U_{\infty}$ is the free-stream velocity, and $\nu_D$ is the kinematic viscosity of the fluid.

In the current simulations, we employ O-type flow domain $\Omega$ of the flow past a circular cylinder. The outer boundary of the domain is denoted by $d\Omega _1$, while the surface of the cylinder is represented by $d\Omega _2$. The boundary conditions on $d\Omega _1$  and $d\Omega _2$ are defined as follows:
\begin{equation}
\mathcal{L}_\mathnormal{BC}\;[\bf{u}]=\mathbf{B}\mathnormal{_C}\bf{(x)},
\end{equation}
where $\mathcal{L}_\mathnormal{BC}$ is a linear operator and $\mathbf{B}\mathnormal{_C}\bf{(x)}$ has a prescribed value. No-slip and no-penetration boundary conditions are applied on the surface of circular cylinder. Dirichlet boundary condition $\mathbf{u}=(u,v,w)=(1,0,0)$ is employed at the inflow, whereas Neumann boundary condition $\partial_n \mathbf{u}=0$ is used at the outflow of the computational domain.

\begin{figure}[htb]
\centering
\subfigure[]{\includegraphics[angle=0,width=0.49\linewidth]{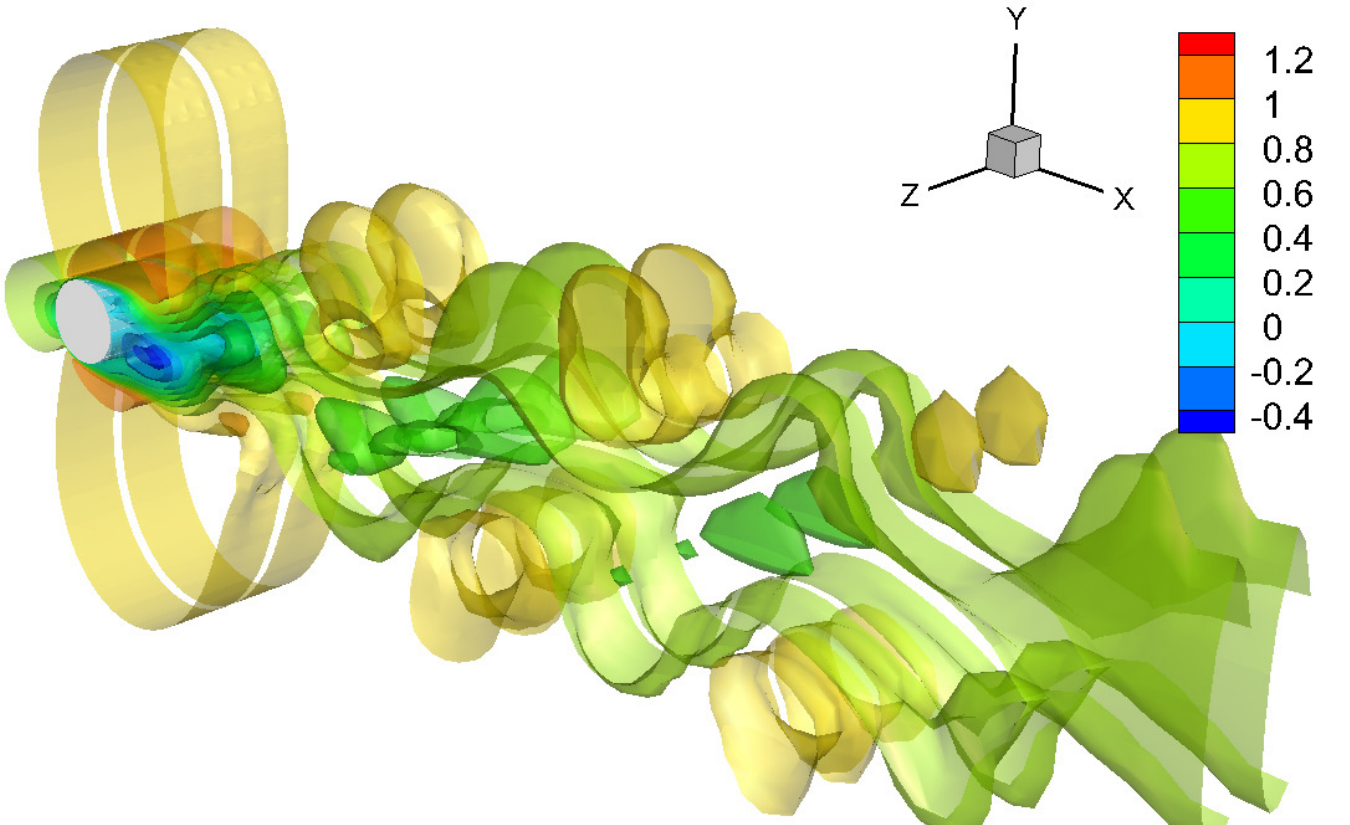}}
\subfigure[]{\includegraphics[angle=0,width=0.49\linewidth]{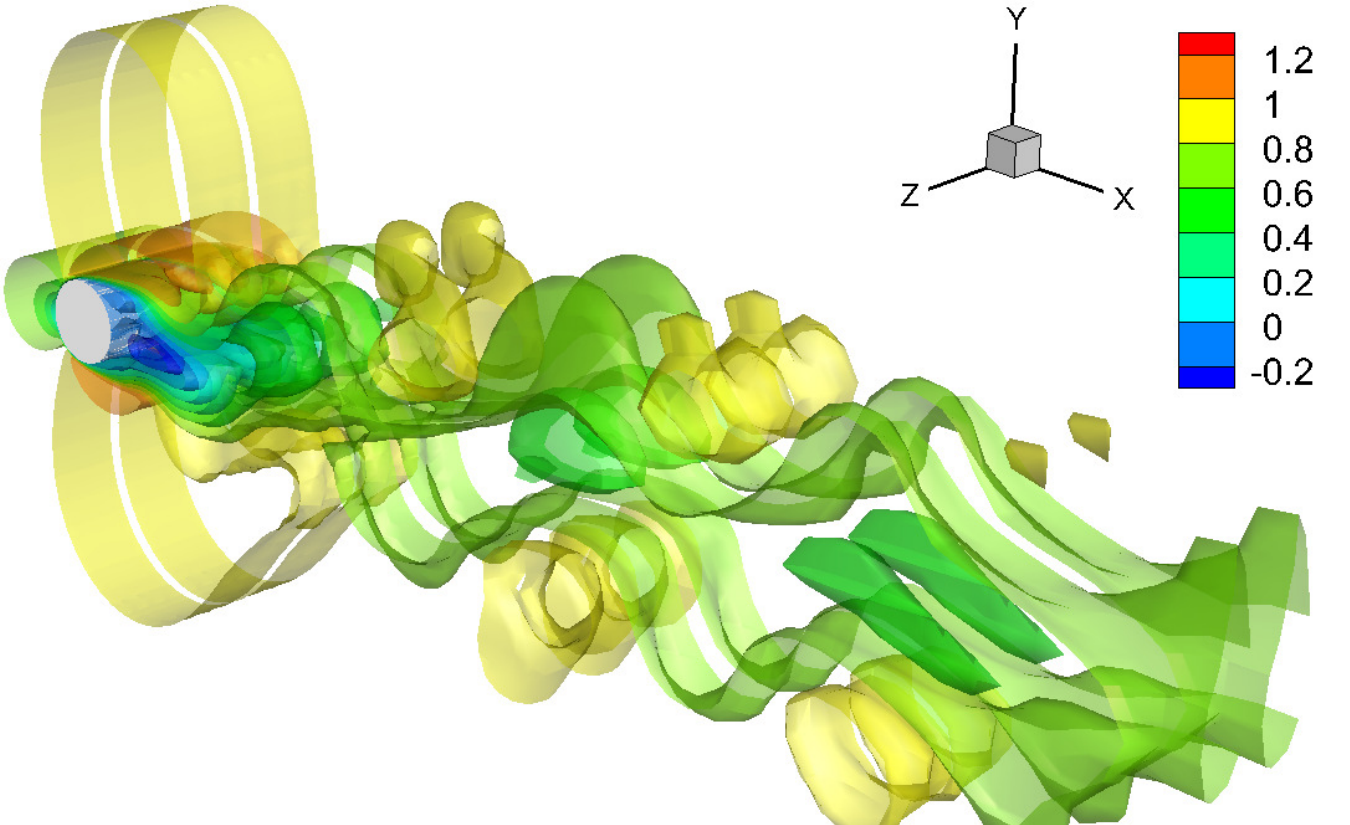}}
\subfigure[]{\includegraphics[angle=0,width=0.49\linewidth]{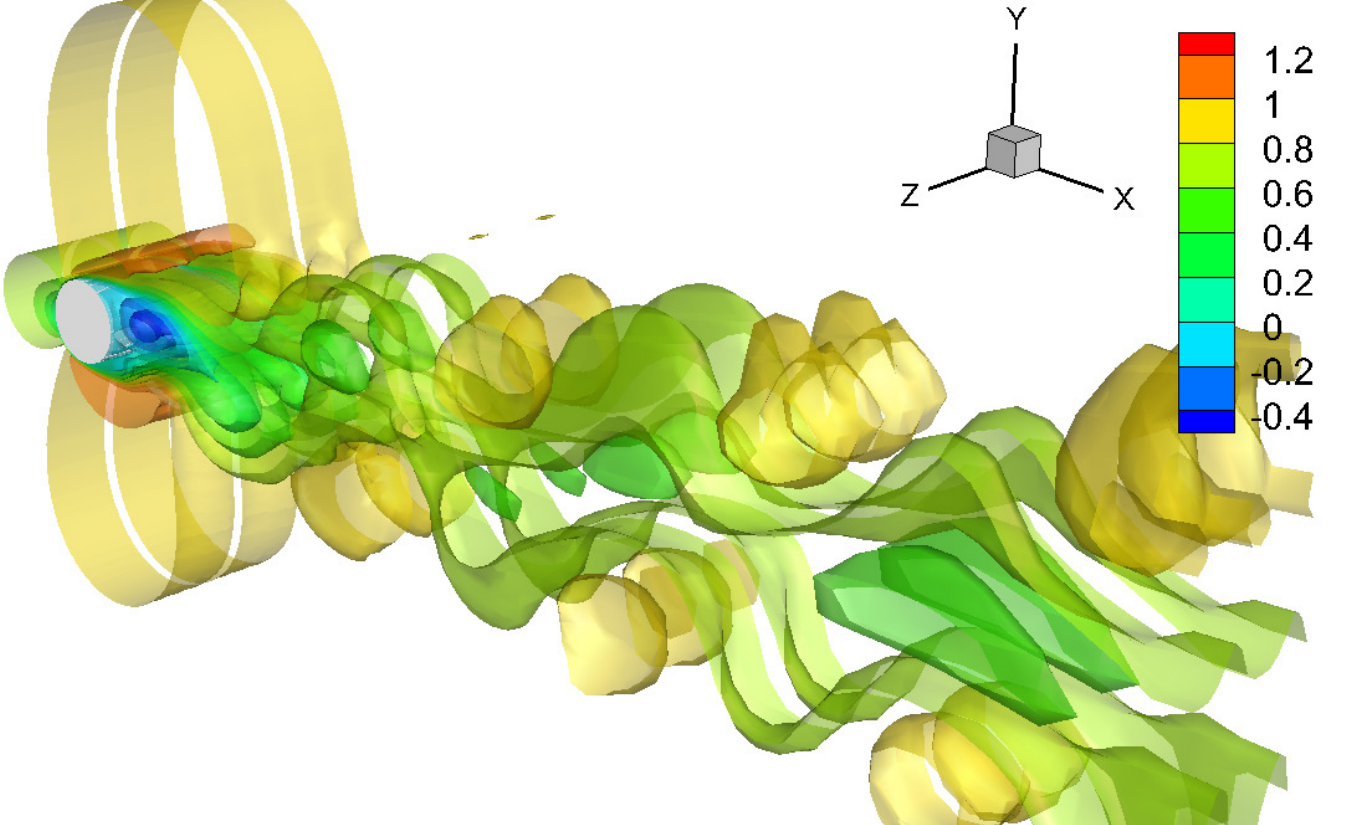}}
\subfigure[]{\includegraphics[angle=0,width=0.49\linewidth]{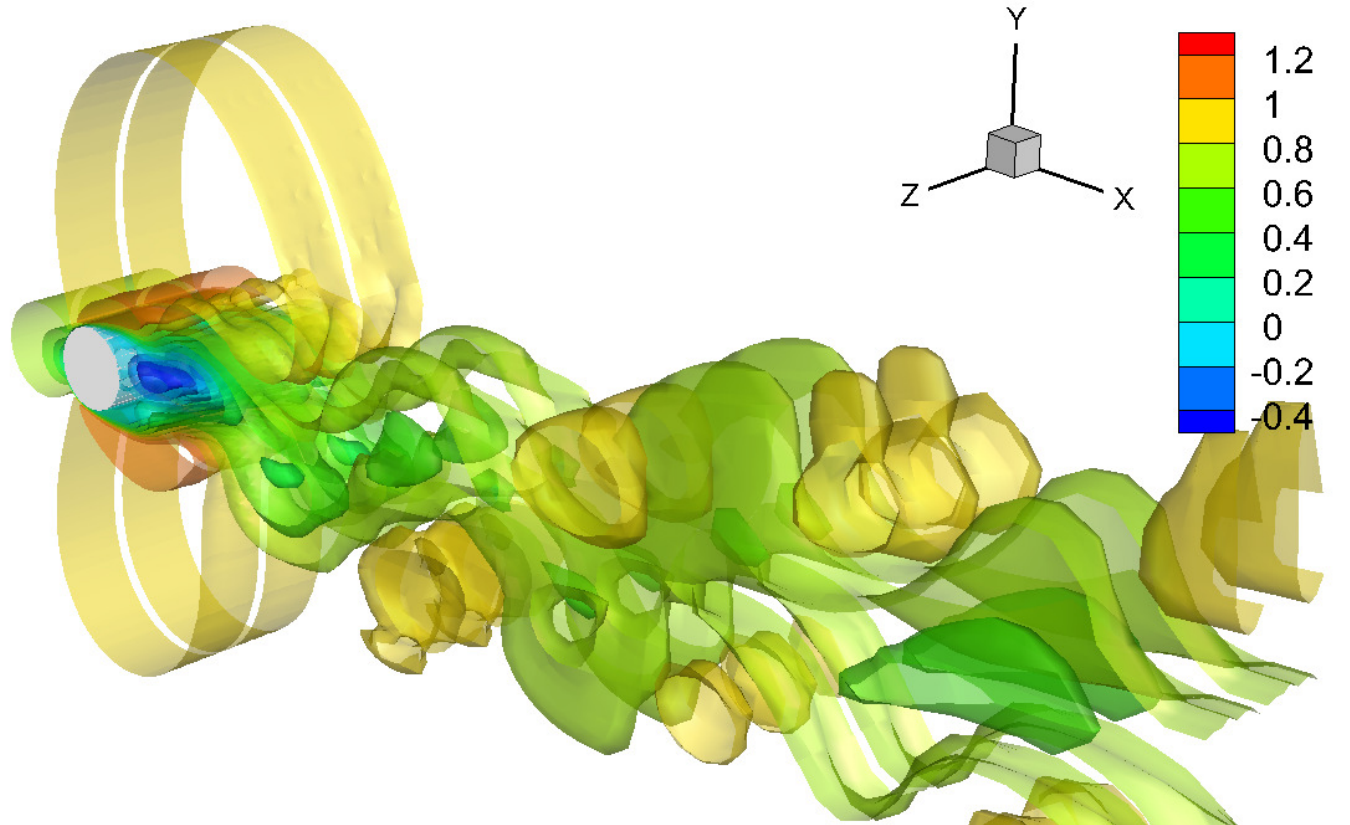}}
\caption{Iso-surfaces of the streamwise velocity field at $\rm Re$ = 1000 at (a) $t = \frac{1}{4} T_s$, (b) $t = \frac{1}{2} T_s$, (c) $t = \frac{3}{4} T_s$, and (d) $t = T_s$.}
\label{fig:streamwise_snapshots}
\end{figure}

We simulate the flow field past a three-dimensional circular cylinder at $\rm{Re}=1000$.
The domain size is 15D with a non-dimensional spanwise length of 2D. Using the polar coordinates ($r, \theta, z$), the domain is divided into $144\times192\times16$ grid points. Thus, the PDE in equation (\ref{eqn:NSE}) with infinite dimensions is transformed into a finite dimensional problem. To validate the results, we compute the mean drag coefficient and the Strouhal number $\mbox{St} = f L/U_{\infty}$. Table \ref{tab:3DNavier} shows that the mean drag coefficient $\bar{C}_D$ and Strouhal number are in a good agreement with the experimental and numerical results in literature  \cite{norberg1994experimental, evangelinos1999dynamics}. For more details on discretization, validation, and verification study, the readers are referred to Akhtar et al. \cite{Akhtar2008C}. Figure \ref{fig:streamwise_snapshots} shows the iso-surfaces of streamwise velocity at $\rm{Re}=$1000 at $t = \frac{1}{4} T_s$, $t = \frac{1}{2} T_s$,  $t = \frac{3}{4} T_s$, and $t = T_s$ where $T_s$ is the vortex shedding time period. It is observed that the flow is turbulent in the wake.

\begin{table}
\centering
\caption{Validation study at $\rm{Re}=1000$.}
\begin{tabular}{lcc}
\addlinespace[2ex]
\toprule
Study  &  $\bar{C}_D$ & St \\
\midrule
Norberg (Experimental) \cite{norberg1994experimental} & 1.00 & 0.210\\
Evangelinos and Karniadakis (3D DNS) \cite{evangelinos1999dynamics}  & 1.02 &  0.202\\
Present         & 1.10     & 0.205\\
\bottomrule
\end{tabular}
\label{tab:3DNavier}
\end{table}

\section{Proper-Orthogonal Decomposition}
To perform the proper orthogonal decomposition, we denote the state variable as $\mathbf{u}(.,t)$, which is evaluated on the Hilbert space $ \mathcal{H}$. For $t\in[0,T]$, $\mathbf{u}(.,t)$ is computed numerically by solving the Navier-Stokes equations (\ref{eqn:NSE}) and capturing the snapshots of the velocity field.  We take the snapshot data of steady state velocity flow field at different time instances $t_1,...,\;t_N\in[0,T] $ and ensemble them in a snapshots matrix form as

\begin{equation}
\centering
{\mathbf{W}:=\textnormal{span}\{ \mathbf{u}(t_1,.),\; \mathbf{u}(t_2,.),\; ...,\;\mathbf{u}(t_N,.)\} } \in \mathbb{R}^{3Q \times N }.
  \label{eq:Snapshots matrix}
\end{equation}
The steady state in a turbulent flow refers to a solution in statistical sense and the dimension of $\mathbf{W}$ is $3Q \times N$.  POD aims to identify low dimensional basis $\{\Phi_1,\;\Phi_2,\;...,\;\Phi_M\}$, where $M<<Q$, that provides an optimal approximation of the snapshots matrix $\mathbf{W}$ as follows:
\begin{equation}
  {\mathop {\min }\limits_{\Phi _j }\frac{1}{N}\sum_{i=1}^N\|\ {\mathbf{u}(t_i,.)} - \sum_{j=1}^M ({\mathbf{u}(t_i,.)},\normalfont\Phi_j(.))_\mathcal{H}\Phi_j(.) \|_\mathcal{H}^2 },
\label{eq:Snapshots approximation}
\end{equation}
where equation (\ref{eq:Snapshots approximation}) is subjected to the following condition: $(\Phi_i,\Phi_j)_{\mathcal{H}}=\delta_{ij}, 1\leq i , j\leq M$,  time step $\delta t=\frac{T}{N-1}$, and the time instances $t_k=k\delta t, \;$ $k=1,2,...,N$. Here, the correlation matrix can be defined as
\begin{equation}
  {\mathbf{K}_{ij}=\frac{1}{N}(\mathbf{u}(t_i,.), \mathbf{u}(t_j,.))}_\mathcal{H},
\label{eq:correlation matrix1}
\end{equation}
where the inner product is defined as $(a_1,b_1)=\int_\Omega {a_1 \cdot b_1}\:d\Omega$. To solve equation (\ref{eq:correlation matrix1}), the following eigenvalue problem is considered
\begin{equation}
  {\bf{K}\alpha=\lambda \alpha},
  \label{eq:eigen_relation}
\end{equation}
where $\bf{K}$ is the snapshots correlation matrix. On the other hand,  $\nu$ and $\lambda$ are the eigenvectors, and the positive eigenvalues, respectively. It can be shown that the solution of equation (\ref{eq:Snapshots approximation}) is given by
\begin{equation}
  {\Phi_k(.)=\frac{1}{\sqrt{\lambda_k}}\sum_{j=1}^N(\alpha_k)_j\mathbf{u}(t_j,.), \quad\quad \normalfont  1\leq k \leq M}.
\label{eq:phi defination}
\end{equation}

We first ensemble the snapshots into the snapshots matrix $\mathbf{W}$ as defined in equation (\ref{eq:Snapshots matrix}). Here, the steady state velocity flow field data ($u, v, w$) is recorded over 15 shedding cycles with 1000 snapshots to develop the snapshots matrix.  We compute the mean velocity components from the snapshots matrix. We then subtract the mean velocity components from the snapshots matrix and determine the POD modes by using the method of snapshots \cite{Sirovich1987A}.

\begin{figure}[htb]
\centering
\subfigure[Energy distribution in each mode]{\includegraphics[angle=0,width=0.49\linewidth]{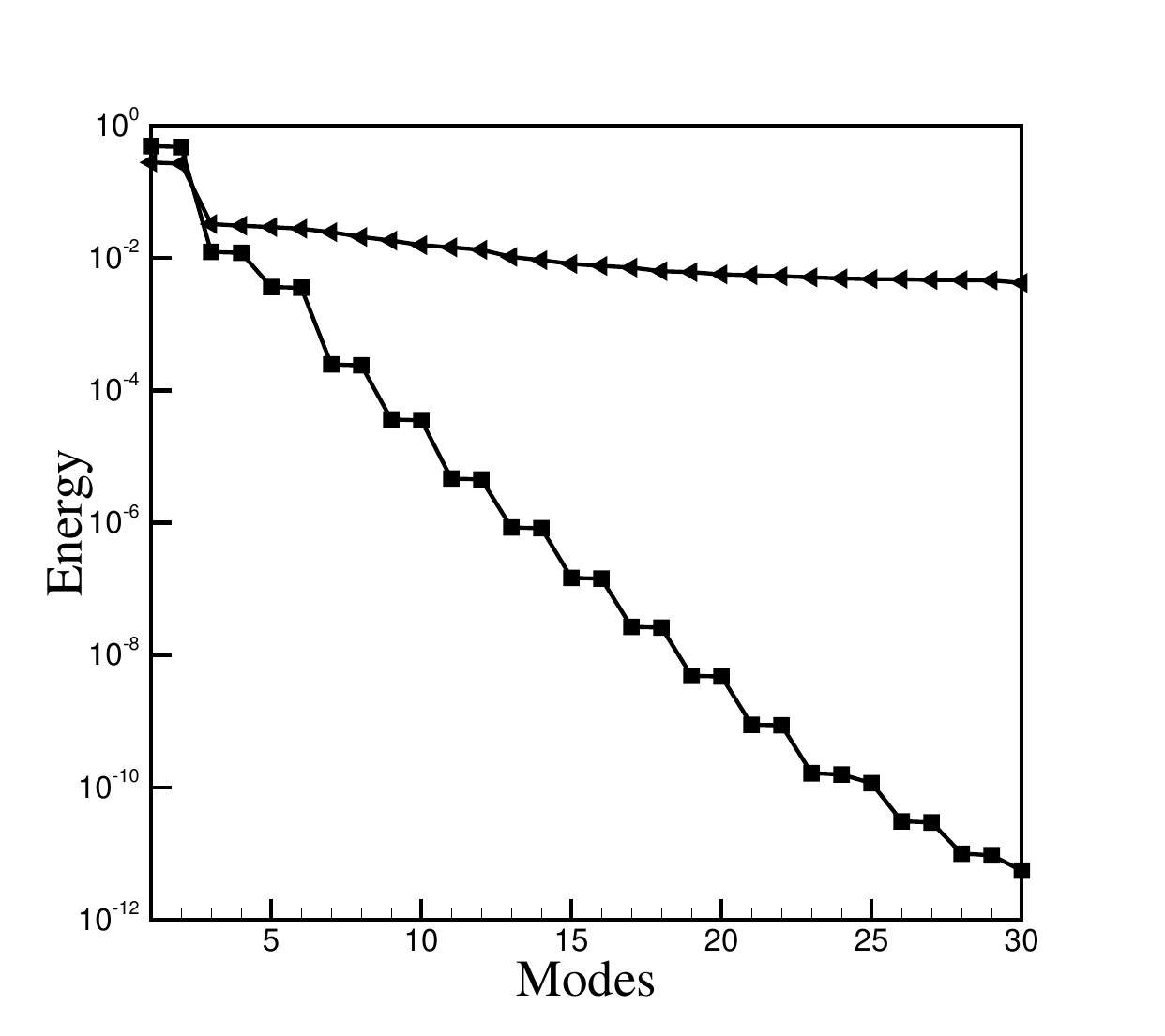}}
\subfigure[Cumulative energy distribution]{\includegraphics[angle=0,width=0.49\linewidth]{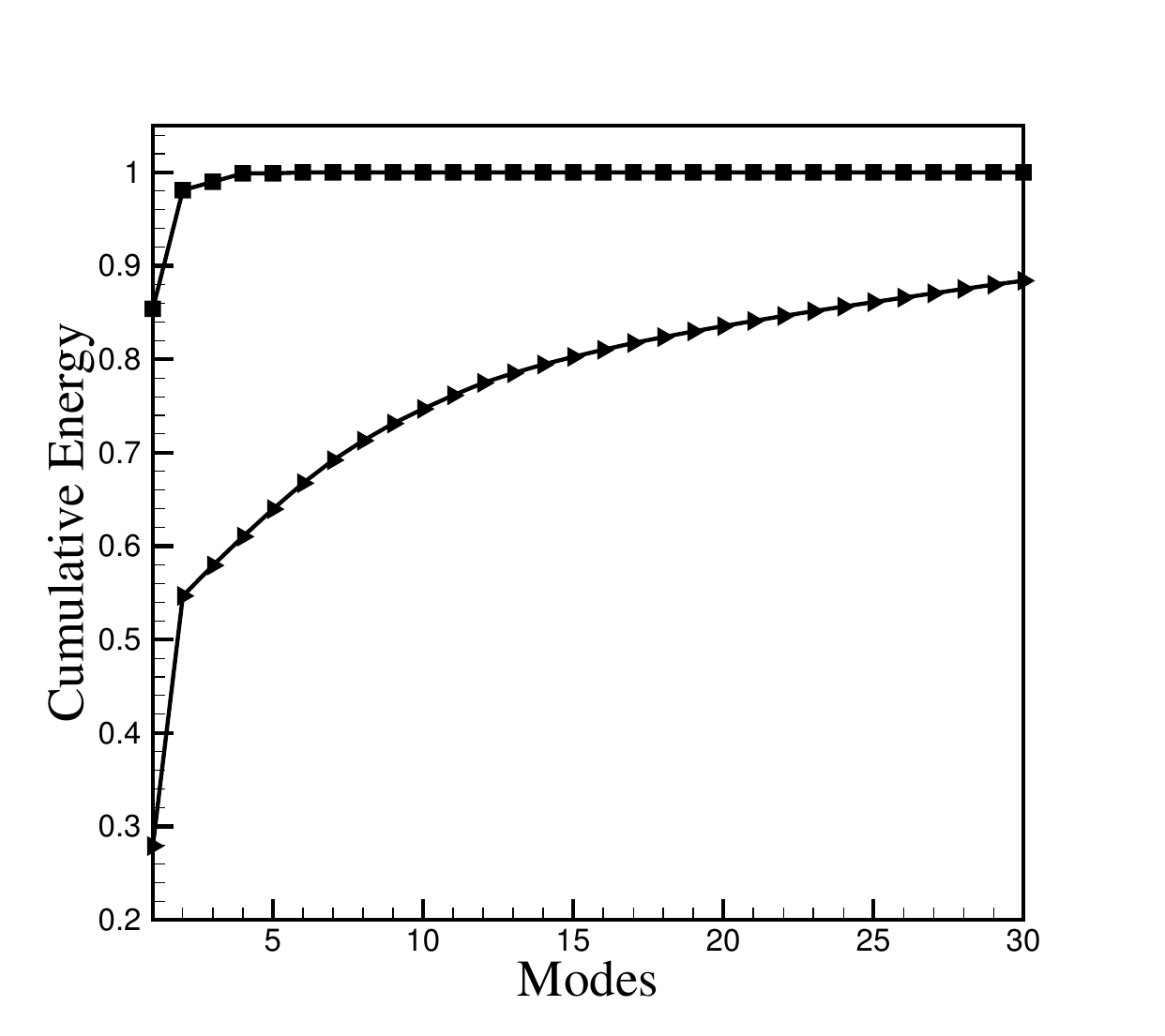}}
\caption{Modal energy distribution for \rm{Re} = 100 (square) and \rm{Re} = 1000 (triangle).}
\label{fig:eigen}
\end{figure}

\begin{figure}[htb]
\centering
\subfigure[]
{\includegraphics[angle=0,width=0.40\linewidth]{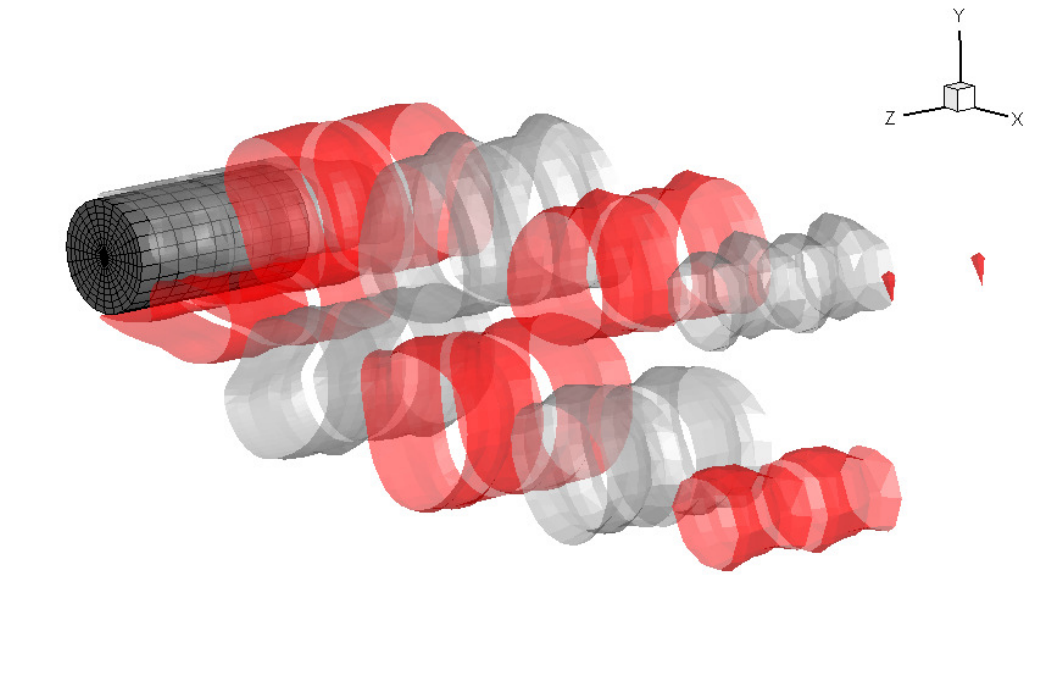}}
\subfigure[]
{\includegraphics[angle=0,width=0.40\linewidth]{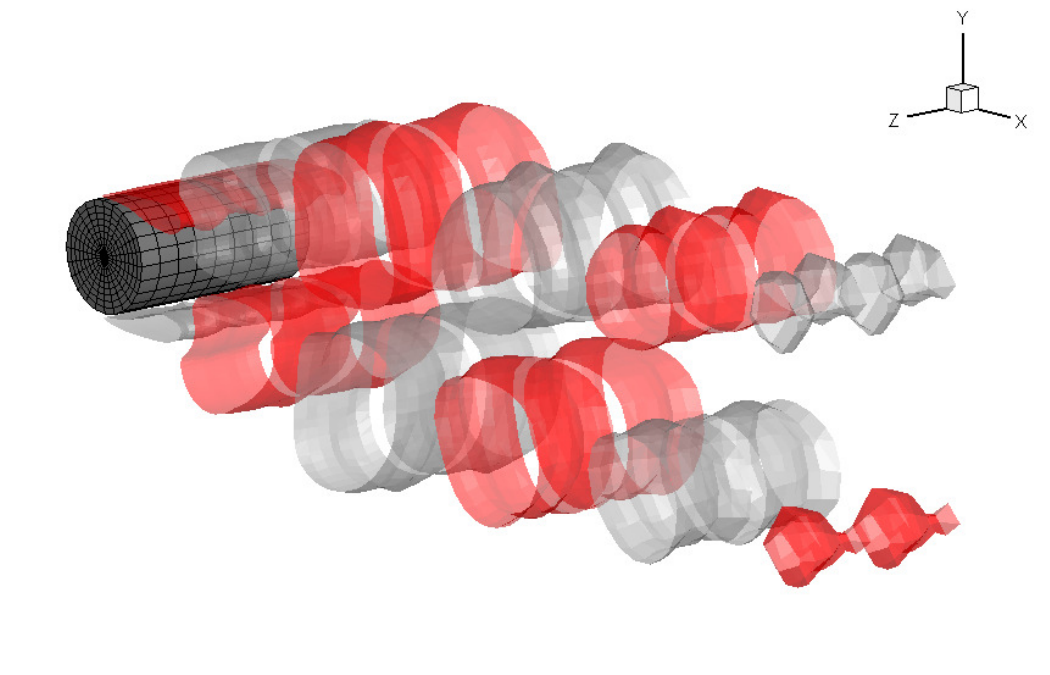}}
\subfigure[]
{\includegraphics[angle=0,width=0.40\linewidth]{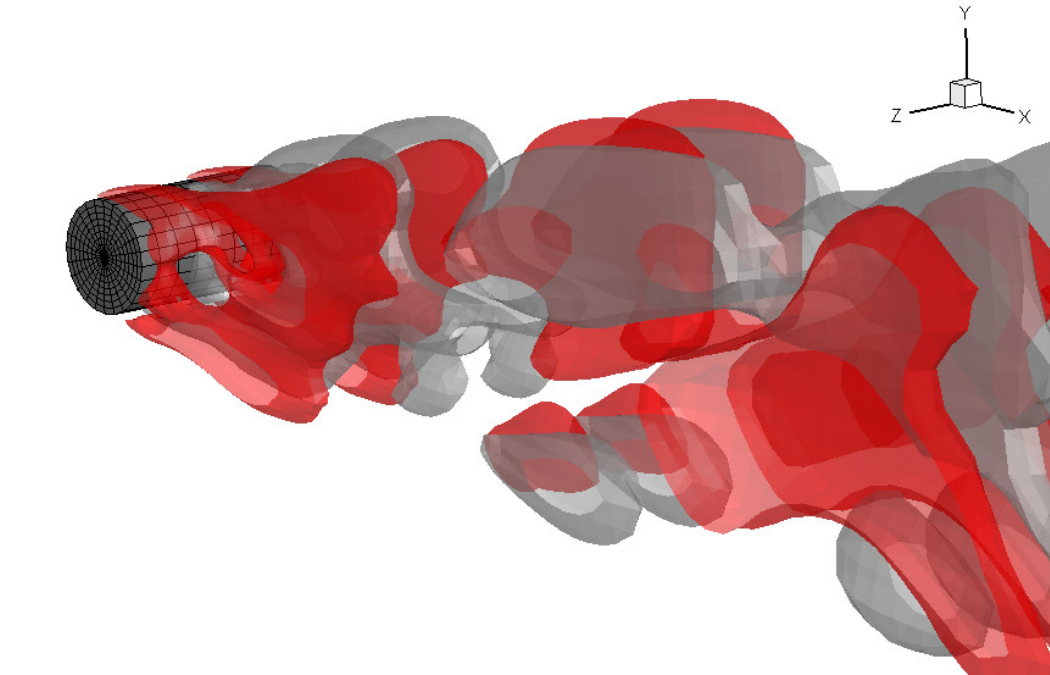}}
\subfigure[]
{\includegraphics[angle=0,width=0.40\linewidth]{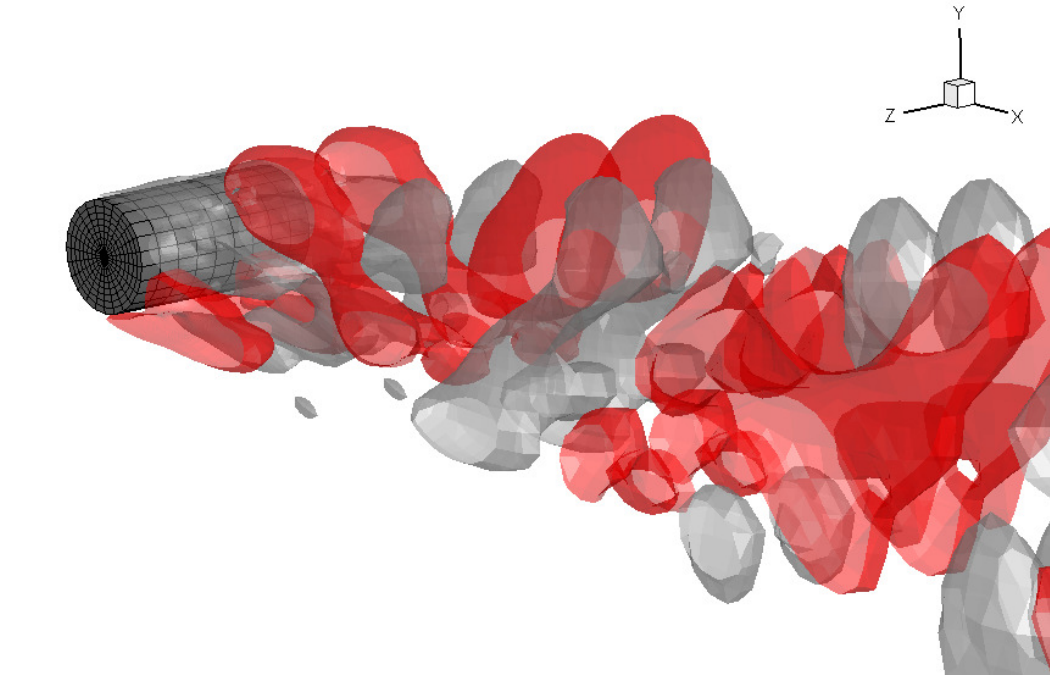}}
\subfigure[]
{\includegraphics[angle=0,width=0.40\linewidth]{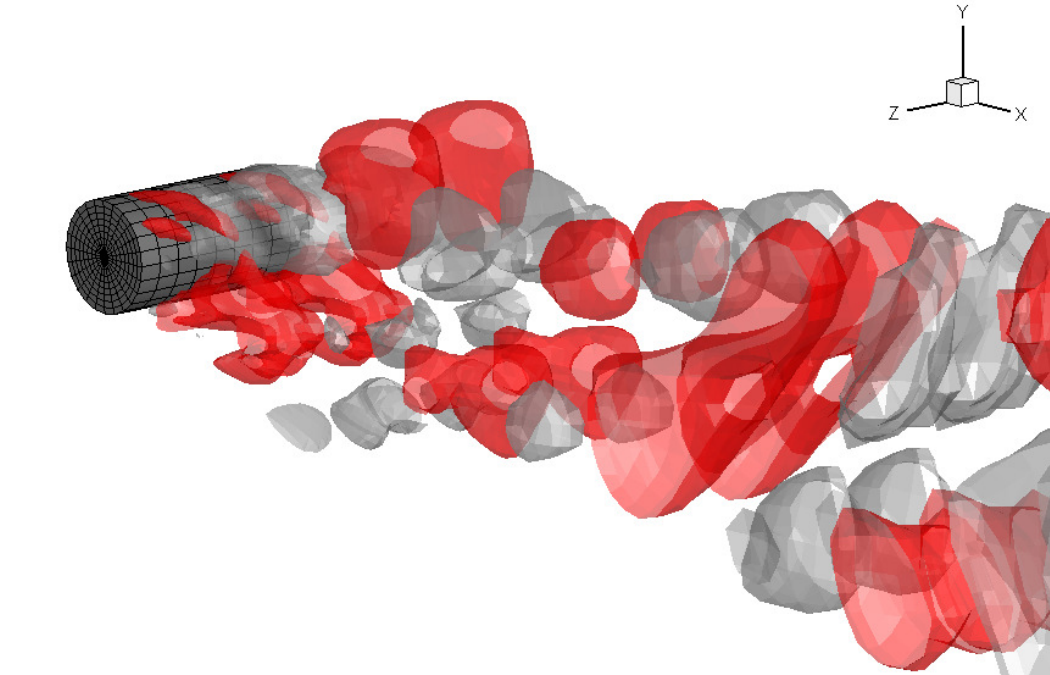}}
\subfigure[]
{\includegraphics[angle=0,width=0.40\linewidth]{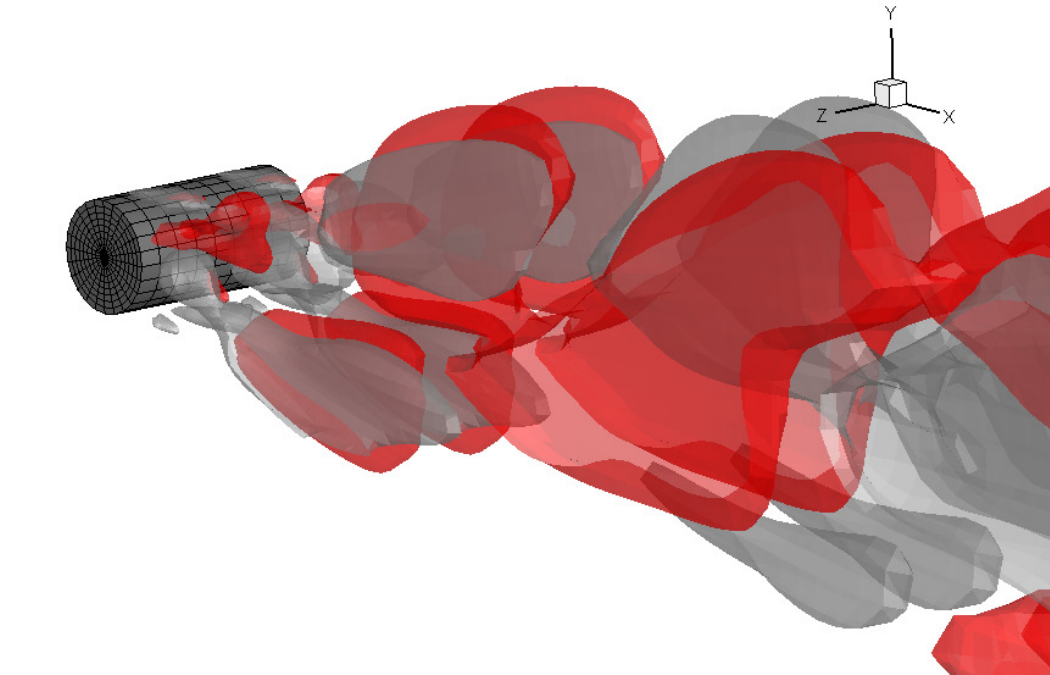}}
\caption{First six POD modes of streamwise velocity component at $\rm{Re} = 1000$.}
\label{fig:3DUmode}
\end{figure}

\begin{figure}[htb]
\centering
\subfigure[]
{\includegraphics[angle=0,width=0.40\linewidth]{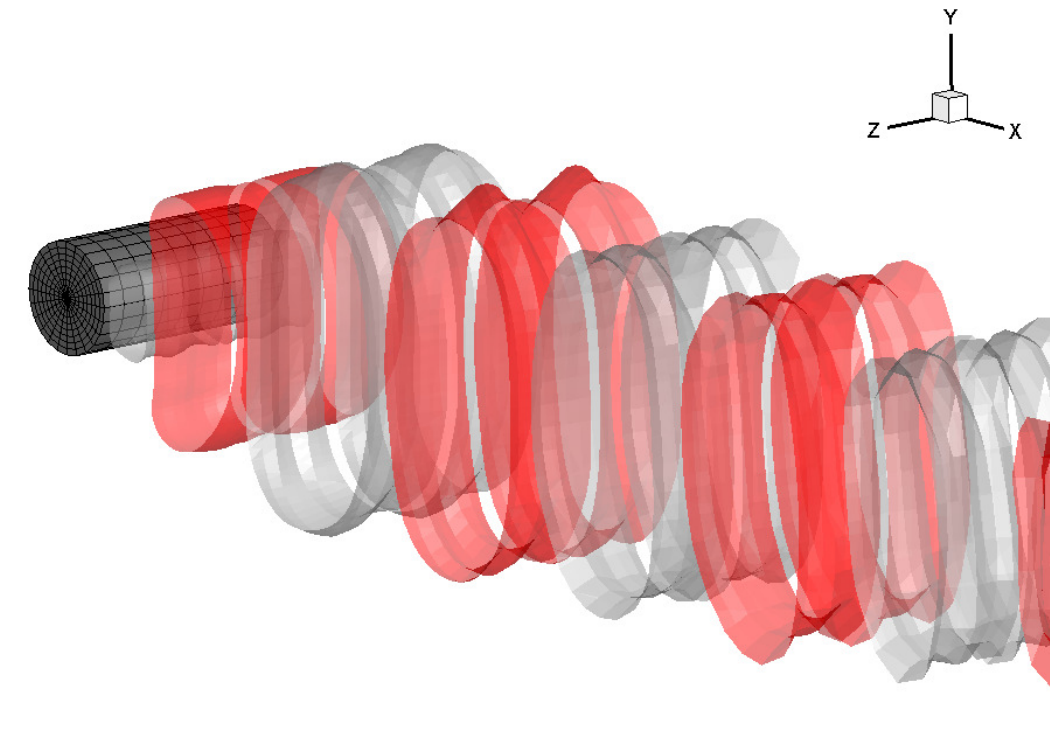}}
\subfigure[]
{\includegraphics[angle=0,width=0.40\linewidth]{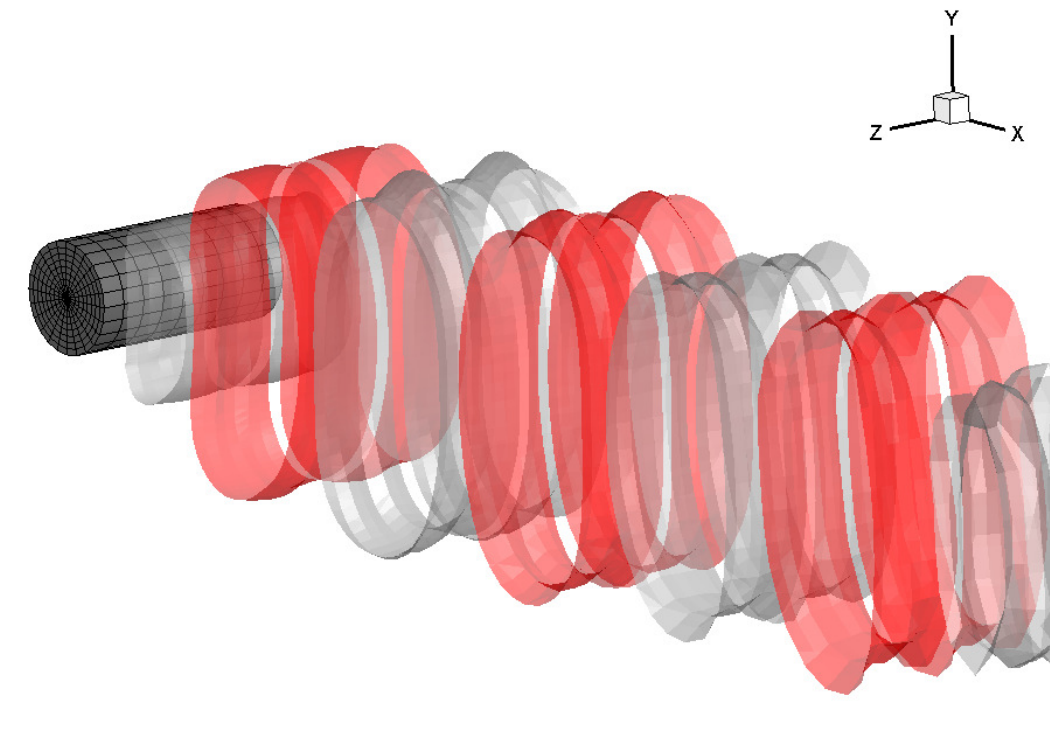}}
\subfigure[]
{\includegraphics[angle=0,width=0.40\linewidth]{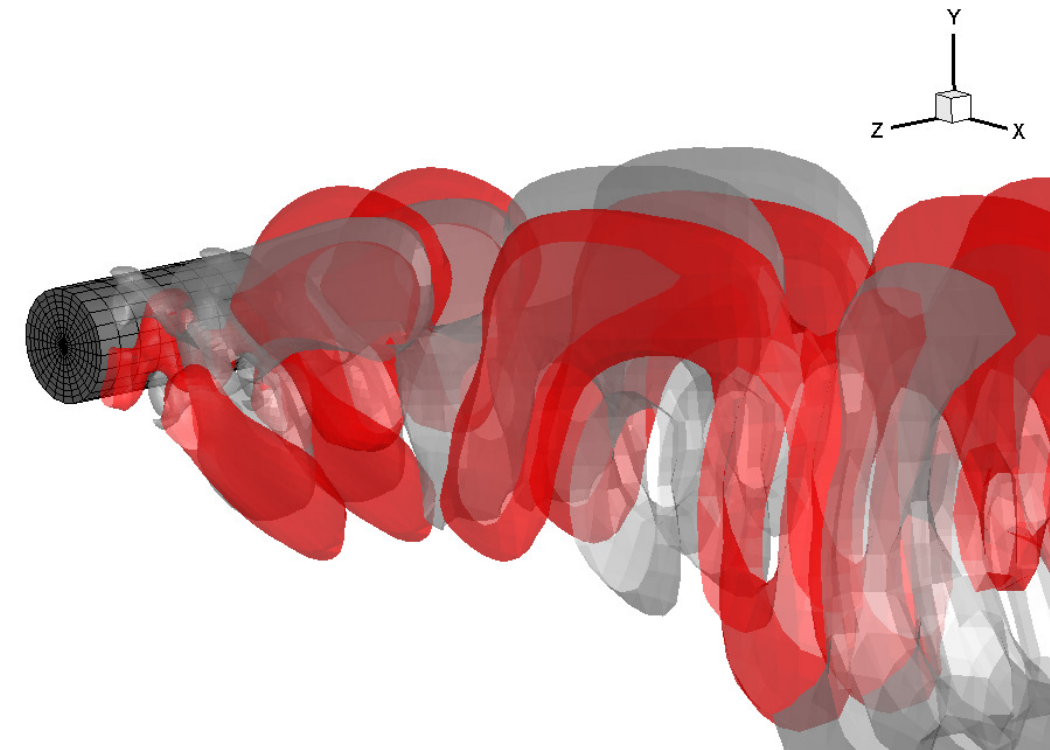}}
\subfigure[]
{\includegraphics[angle=0,width=0.40\linewidth]{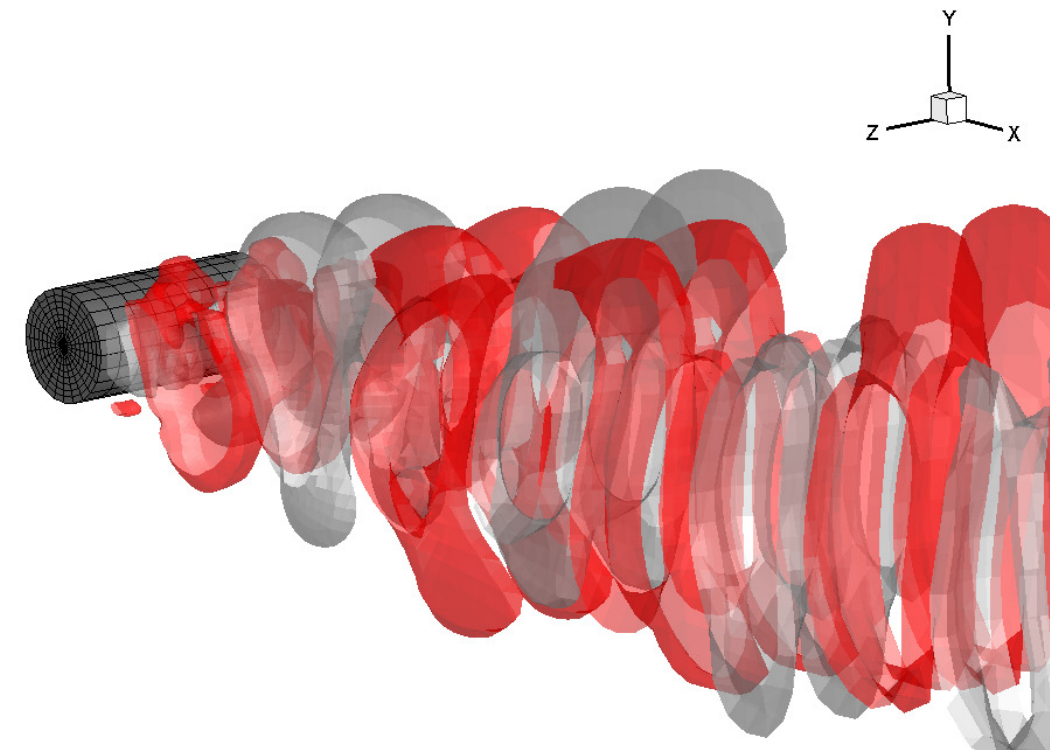}}
\subfigure[]
{\includegraphics[angle=0,width=0.40\linewidth]{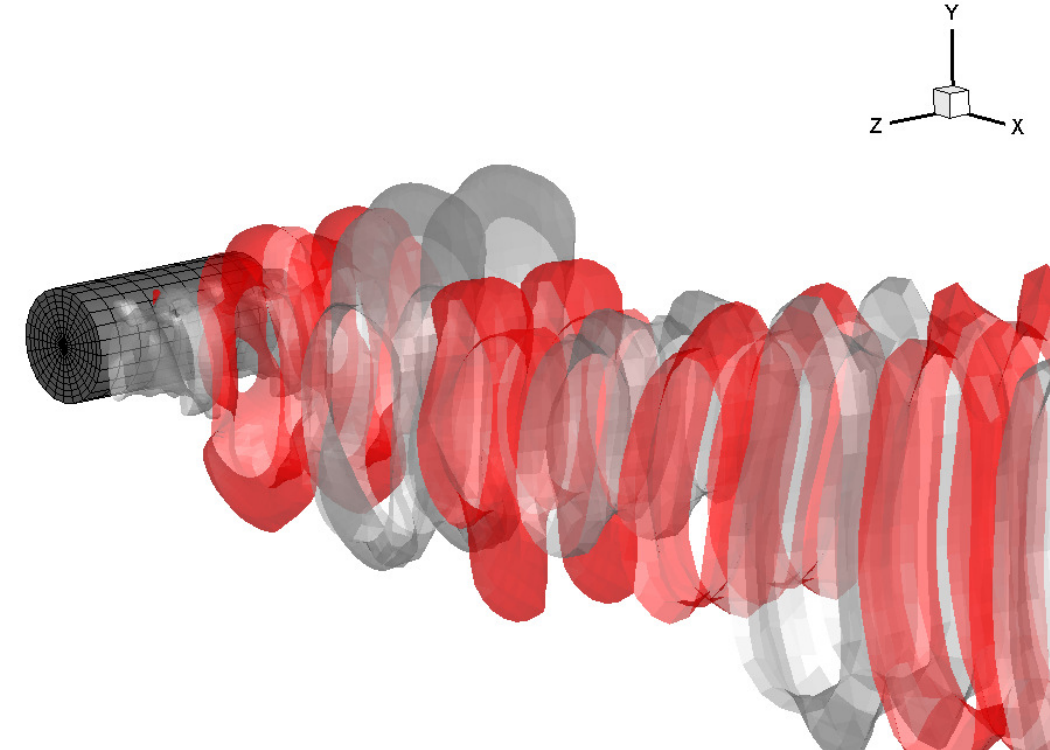}}
\subfigure[]
{\includegraphics[angle=0,width=0.40\linewidth]{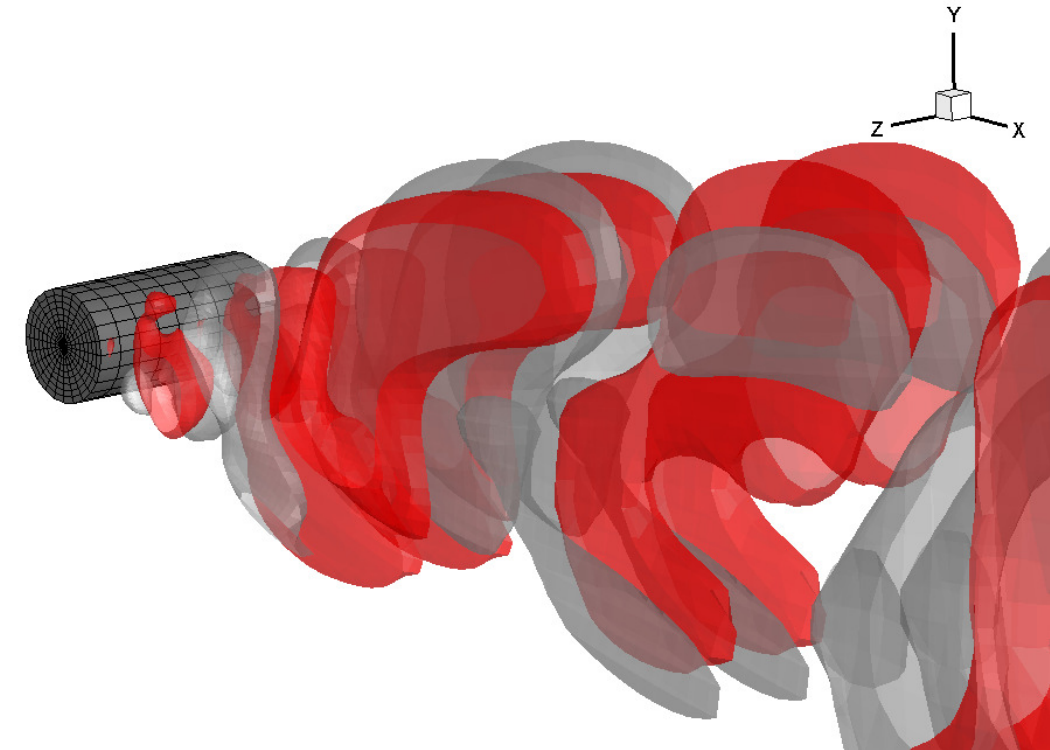}}
\caption{First six POD modes of crossflow velocity component at $\rm{Re} = 1000$.}
\label{fig:3DVmode}
\end{figure}

\begin{figure}[htb]
\centering
\subfigure[]
{\includegraphics[angle=0,width=0.40\linewidth]{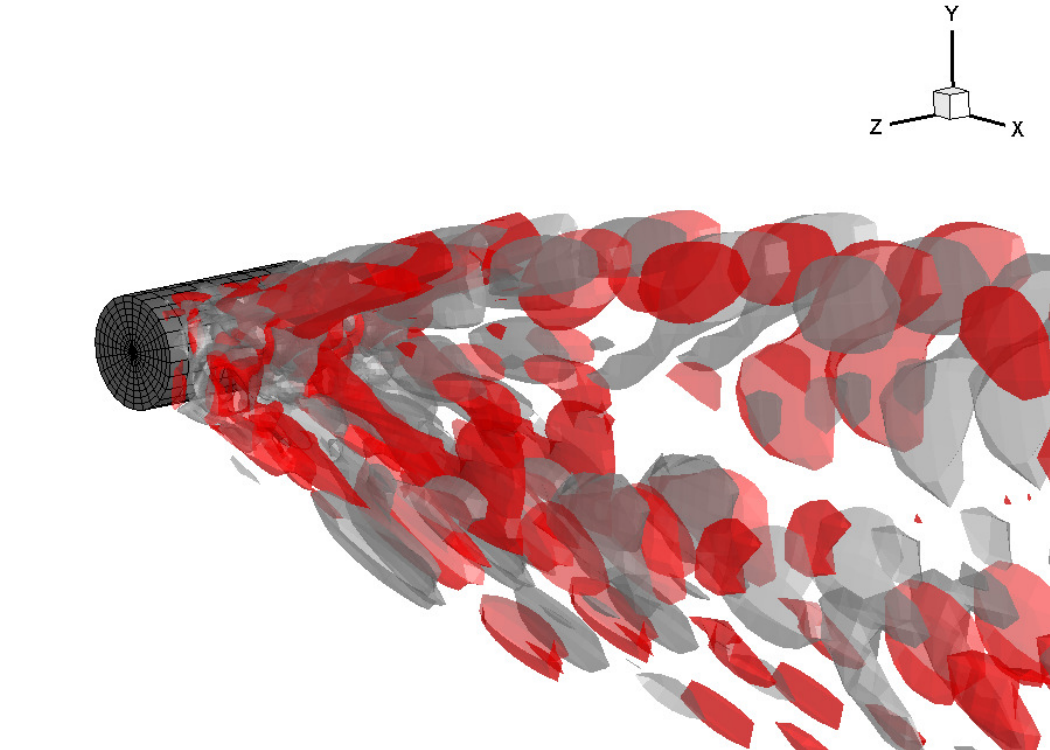}}
\subfigure[]
{\includegraphics[angle=0,width=0.40\linewidth]{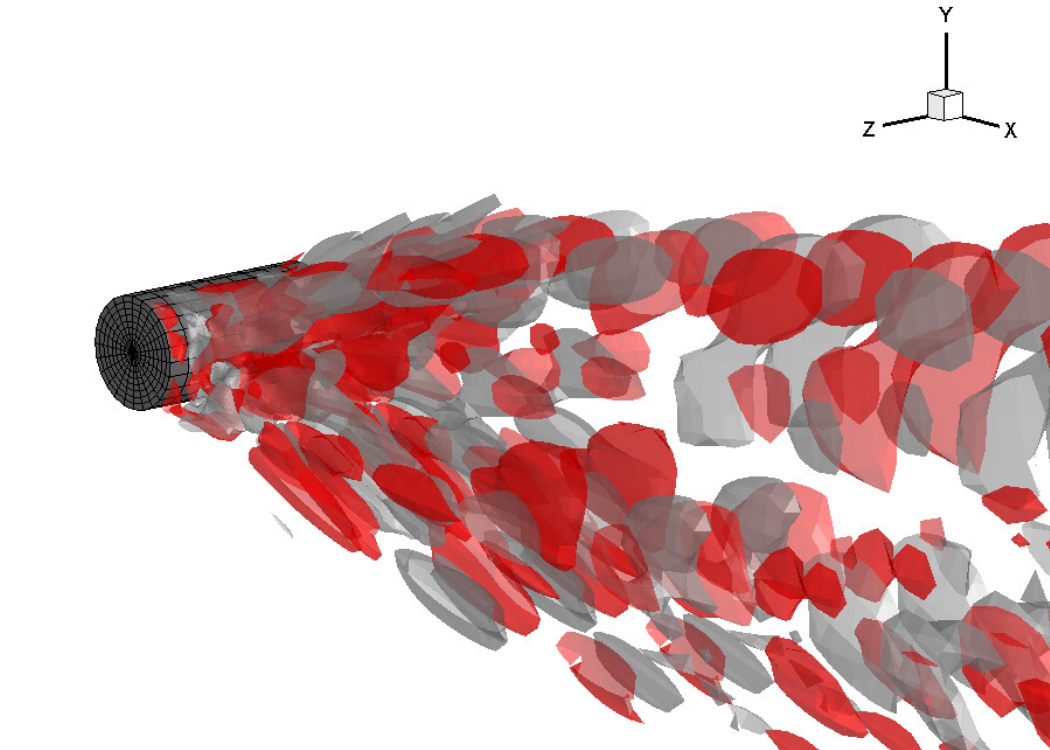}}
\subfigure[]
{\includegraphics[angle=0,width=0.40\linewidth]{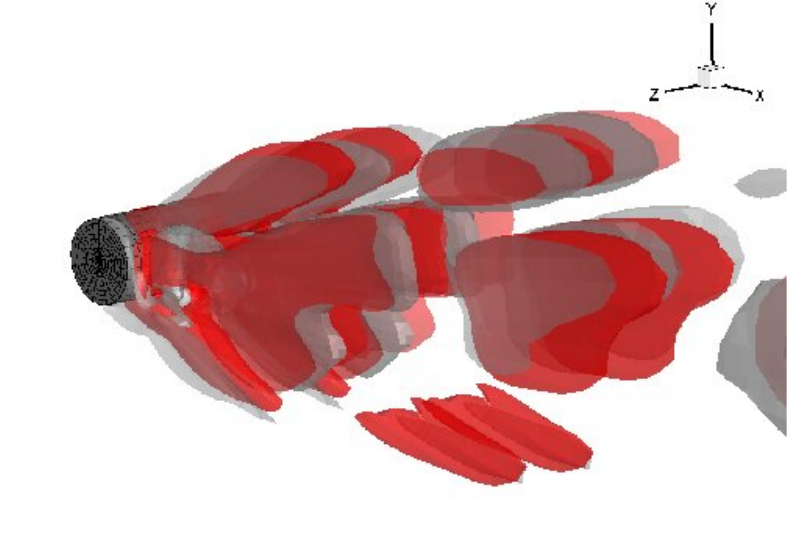}}
\subfigure[]
{\includegraphics[angle=0,width=0.40\linewidth]{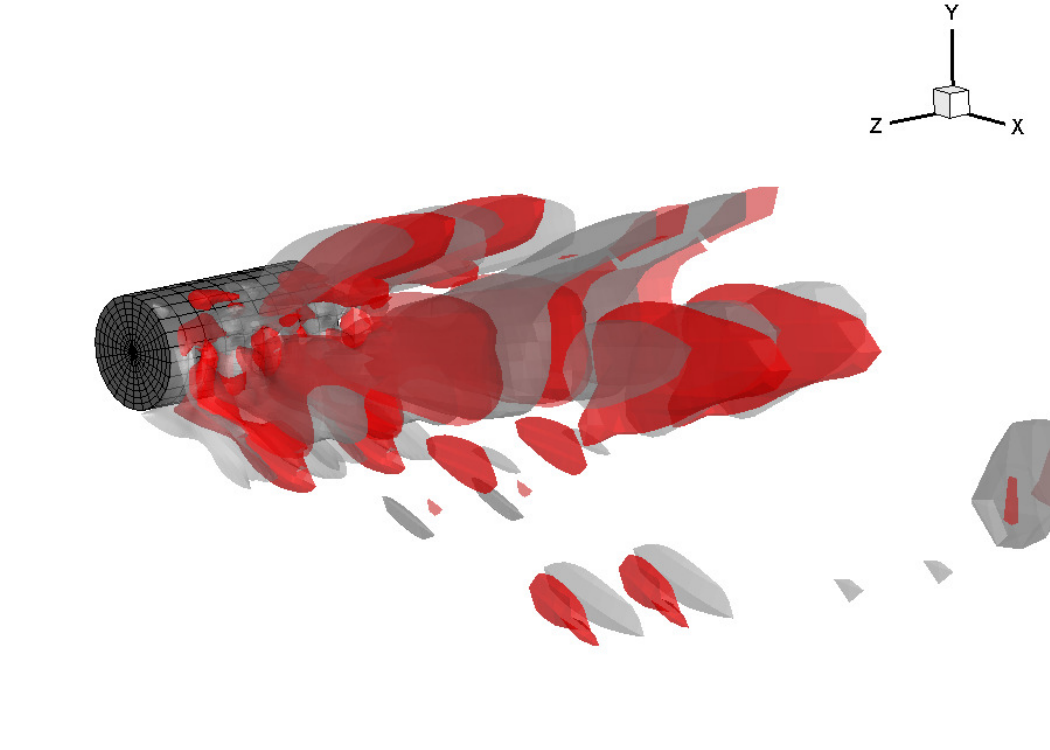}}
\subfigure[]
{\includegraphics[angle=0,width=0.40\linewidth]{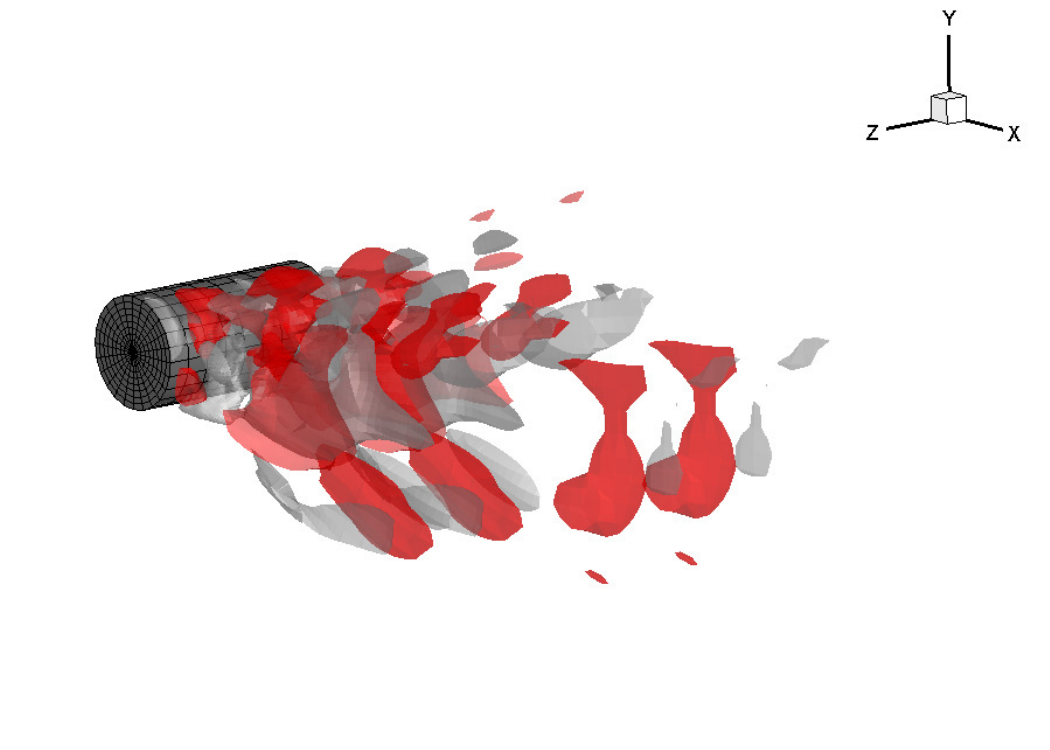}}
\subfigure[]
{\includegraphics[angle=0,width=0.40\linewidth]{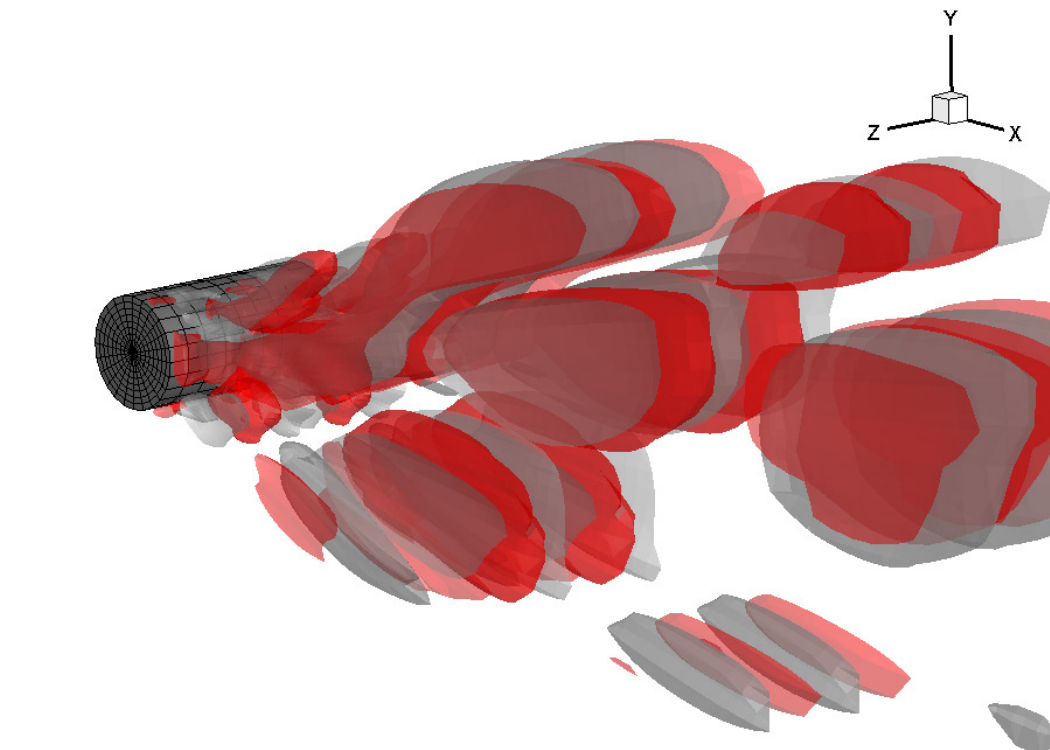}}
\caption{First six POD modes of spanwise velocity component at $\rm{Re} = 1000$.}
\label{fig:3DWmode}
\end{figure}

The eigenvalues ($\lambda_i$) as shown in equation (\ref{eq:eigen_relation})  represent the energy content of the dynamical system for each POD mode.  We plot the first 30 normalized eigenvalues ($\lambda_i/\sum\nolimits_{j = 1}^{N} {\lambda _j}$) as shown in Fig.~\ref{fig:eigen}.  The first two POD modes contain most of the energy content when compared to the remaining POD modes.  The cumulative energy of the modes can be defined as $\sum\nolimits_{i = 1}^M {\lambda _i}/\sum\nolimits_{i = 1}^{N} {\lambda _i}$, which approximates of energy content in each POD mode.  For example, the energy captured by the first 2, 6, 10, and 100 modes is $54.70\%$, $66.70\%$, $74.70\%$, and $98.90\%$ of the dynamical system's total energy, respectively. It's worth emphasizing that the first 100 POD modes in turbulent flows doesn't fully encompass the energy content of higher modes. In contrast, the first 10 modes in a laminar flow field at $\rm{Re}=100$ capture an impressive 99.99\% of the energy content \cite{akhtar2009stability}. Therefore, modeling the physics associated with higher POD modes poses a significant challenge in turbulent flows. Figures \ref{fig:3DUmode}, \ref{fig:3DVmode}, and \ref{fig:3DWmode} present the contour of the first six velocity modes for streamwise, crossflow, and spanwise velocity components, respectively. It is important to note that the effect of third dimensionality is present in spanwise direction of each POD mode.
\FloatBarrier

\FloatBarrier
\FloatBarrier
\FloatBarrier

\begin{figure}[htb]
\centering
\includegraphics[angle=0,width=1.00\linewidth]{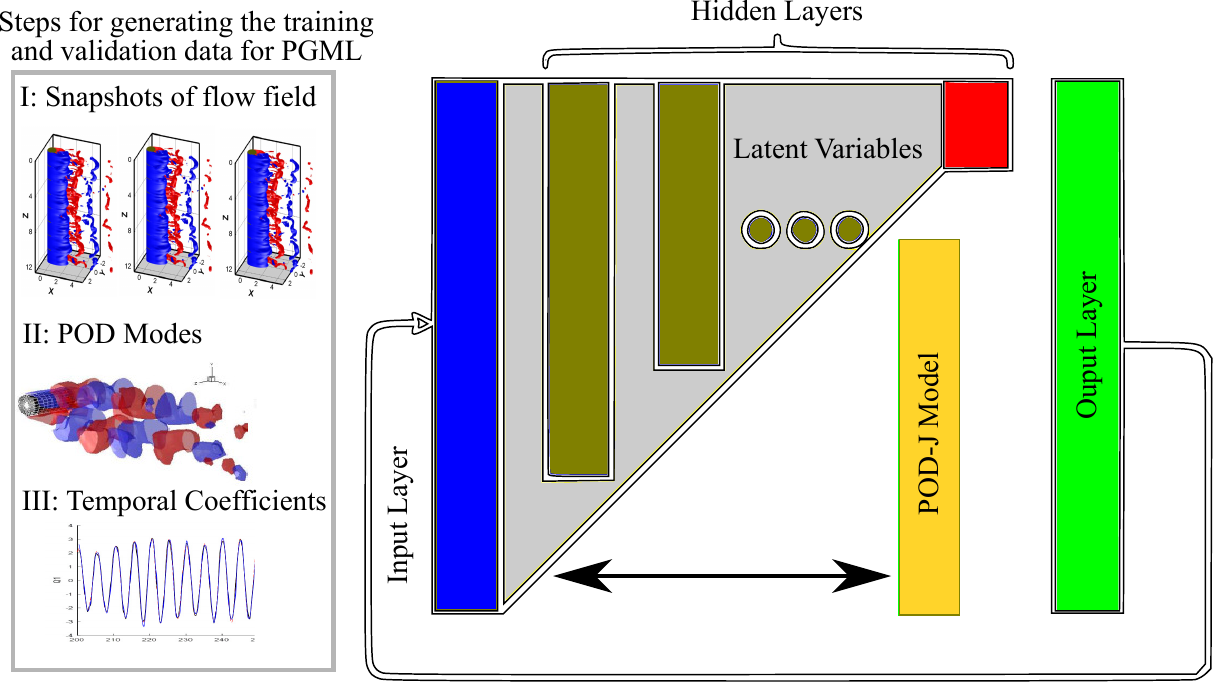}
\caption {Proposed framework for PGML model where the output of POD-J model is concatenated into the middle layers of deep neural networks without breaking the neural network training process.}
\label{fig:Comparison}
\end{figure}

\begin{figure}[htbp]
\centering
\subfigure[]
{\includegraphics[angle=0,width=0.45\linewidth]{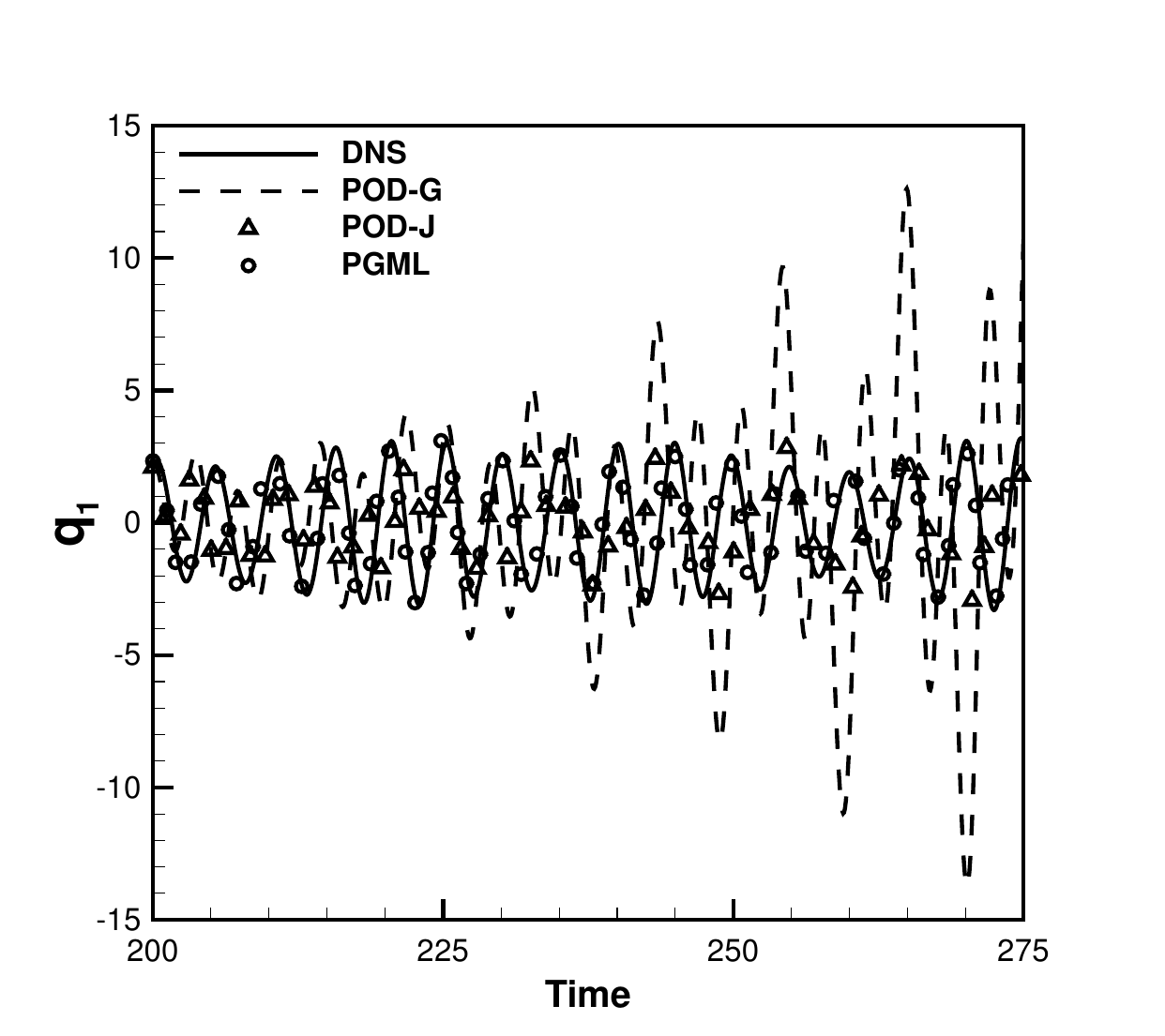}}
\subfigure[]
{\includegraphics[angle=0,width=0.45\linewidth]{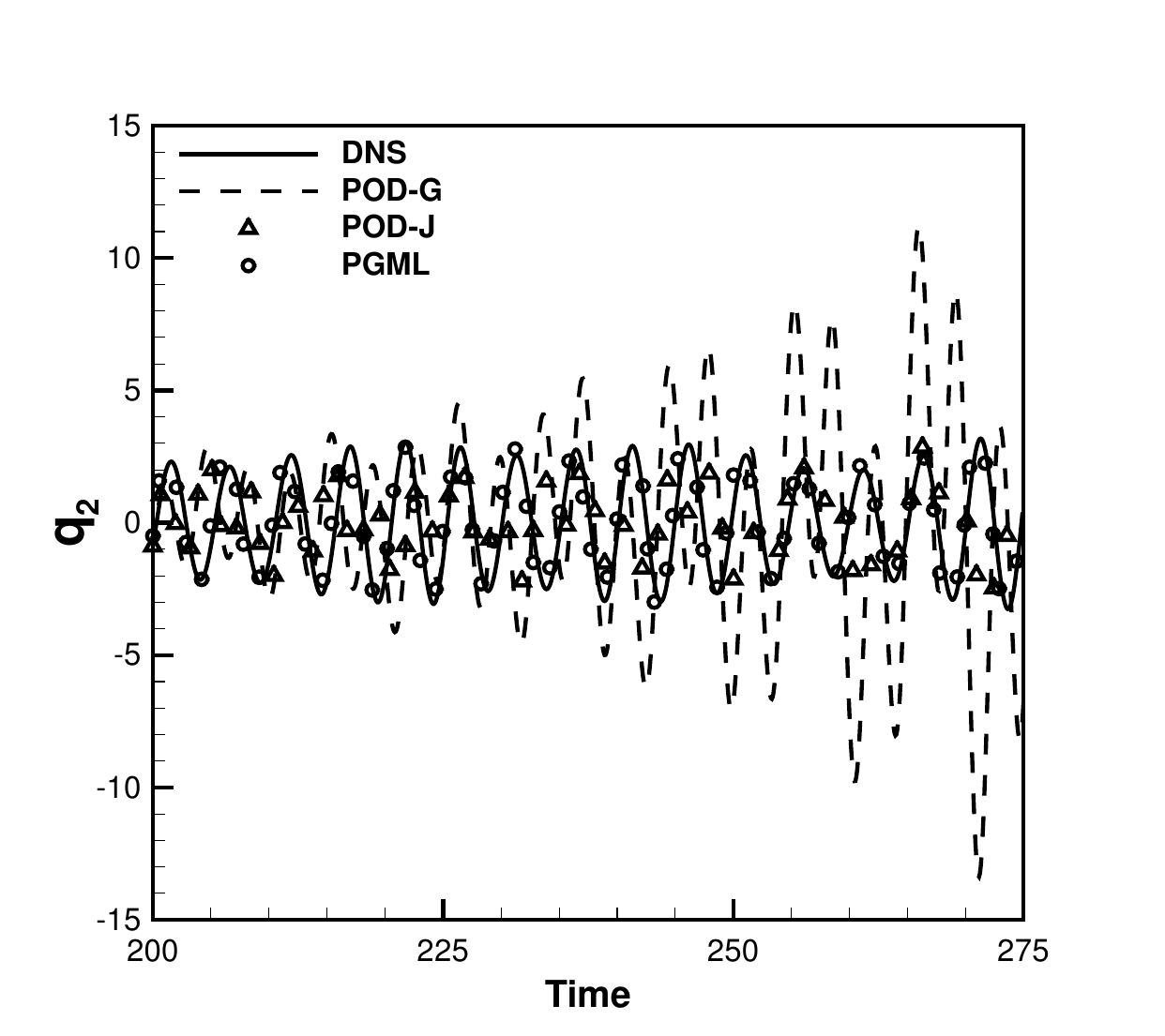}}
\subfigure[]
{\includegraphics[angle=0,width=0.45\linewidth]{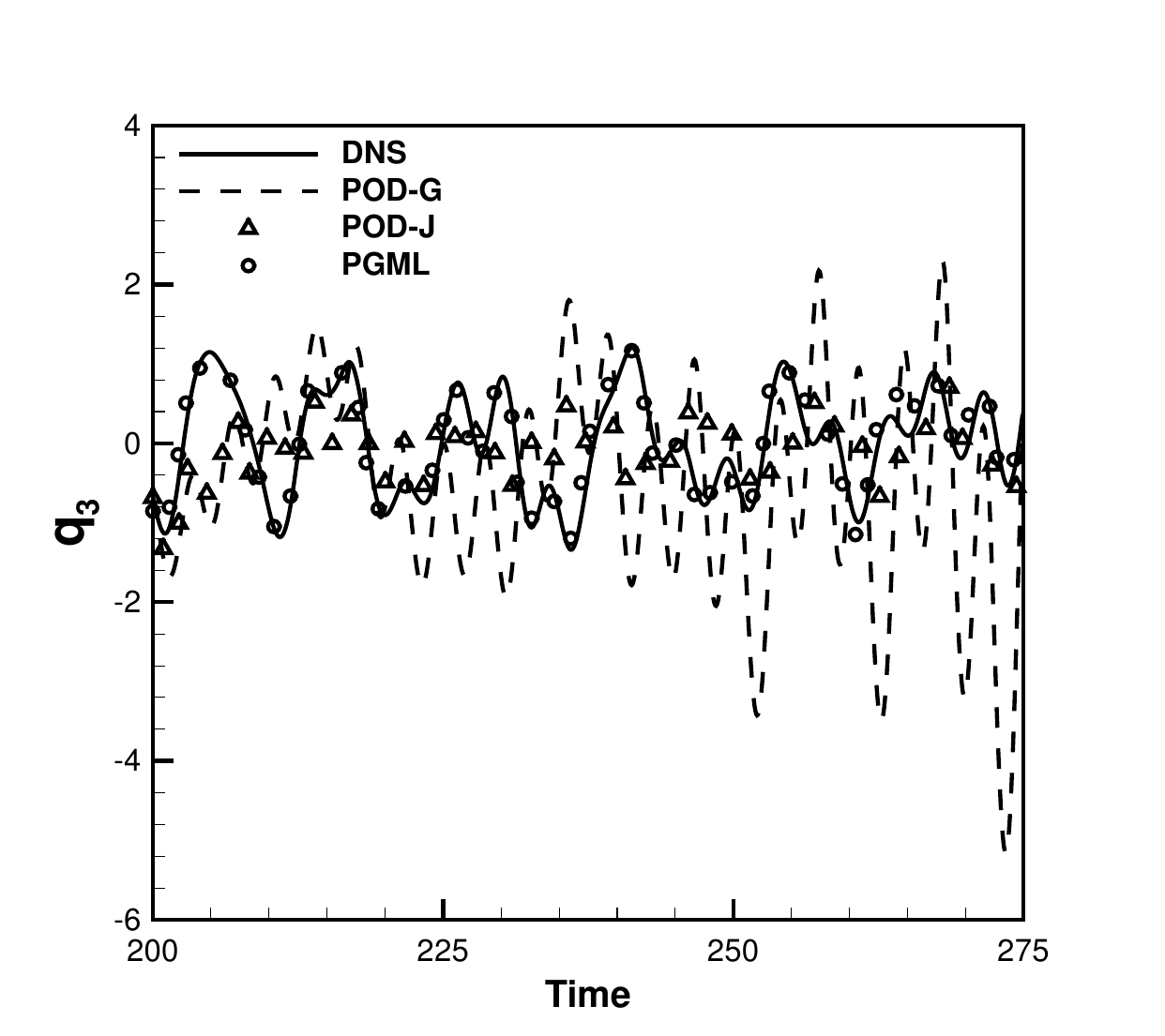}}
\subfigure[]
{\includegraphics[angle=0,width=0.45\linewidth]{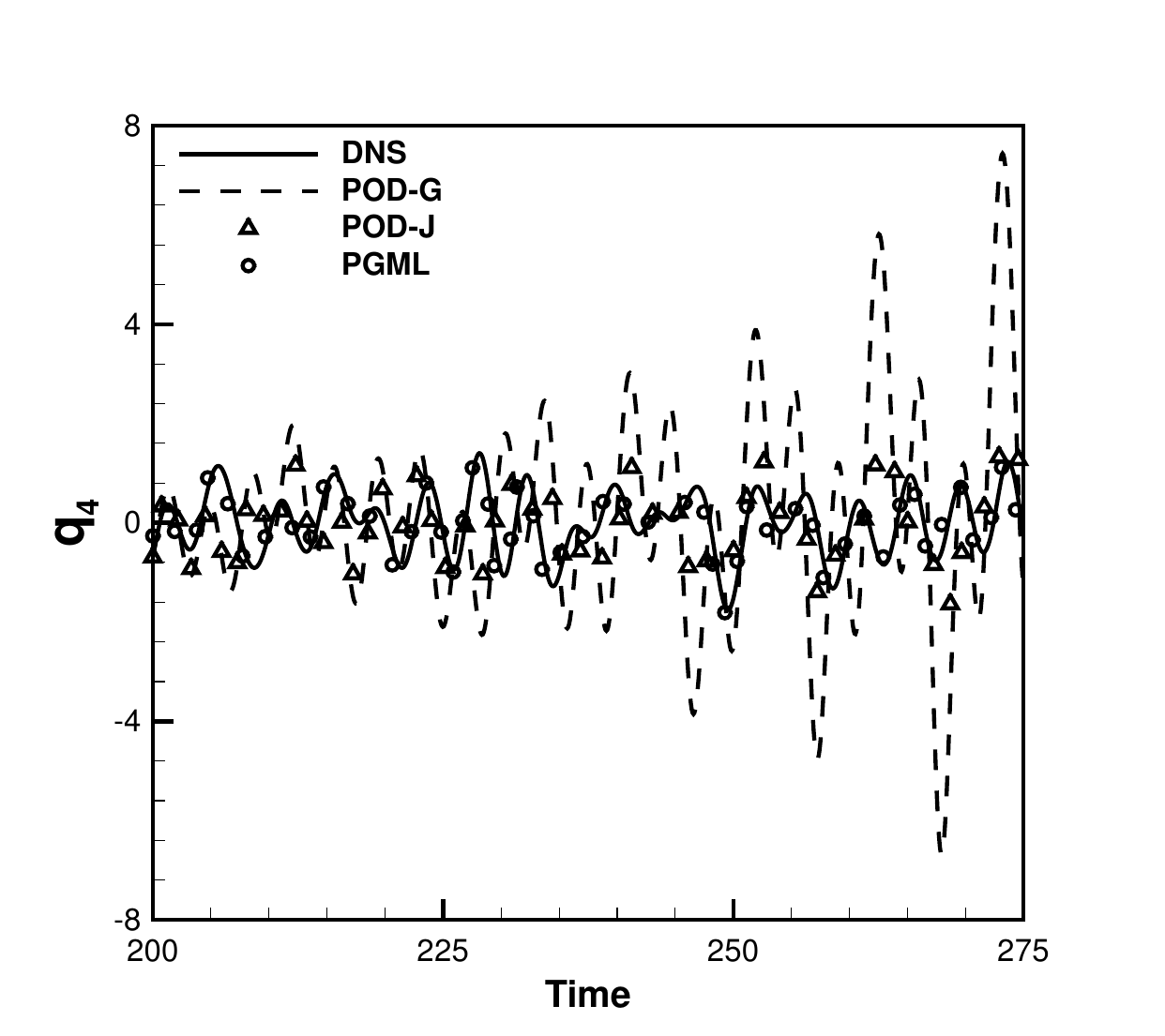}}
\subfigure[]
{\includegraphics[angle=0,width=0.45\linewidth]{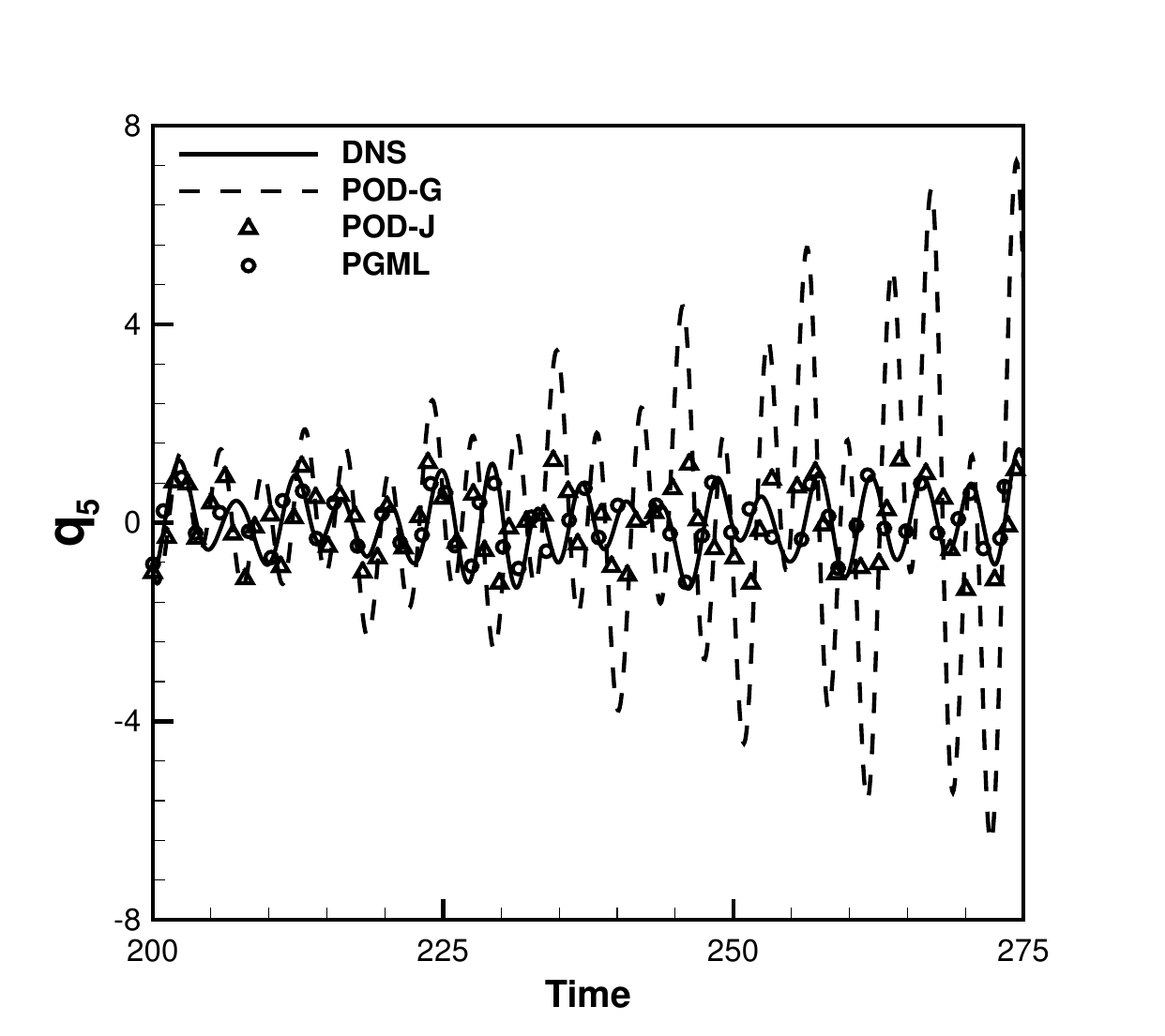}}
\subfigure[]
{\includegraphics[angle=0,width=0.45\linewidth]{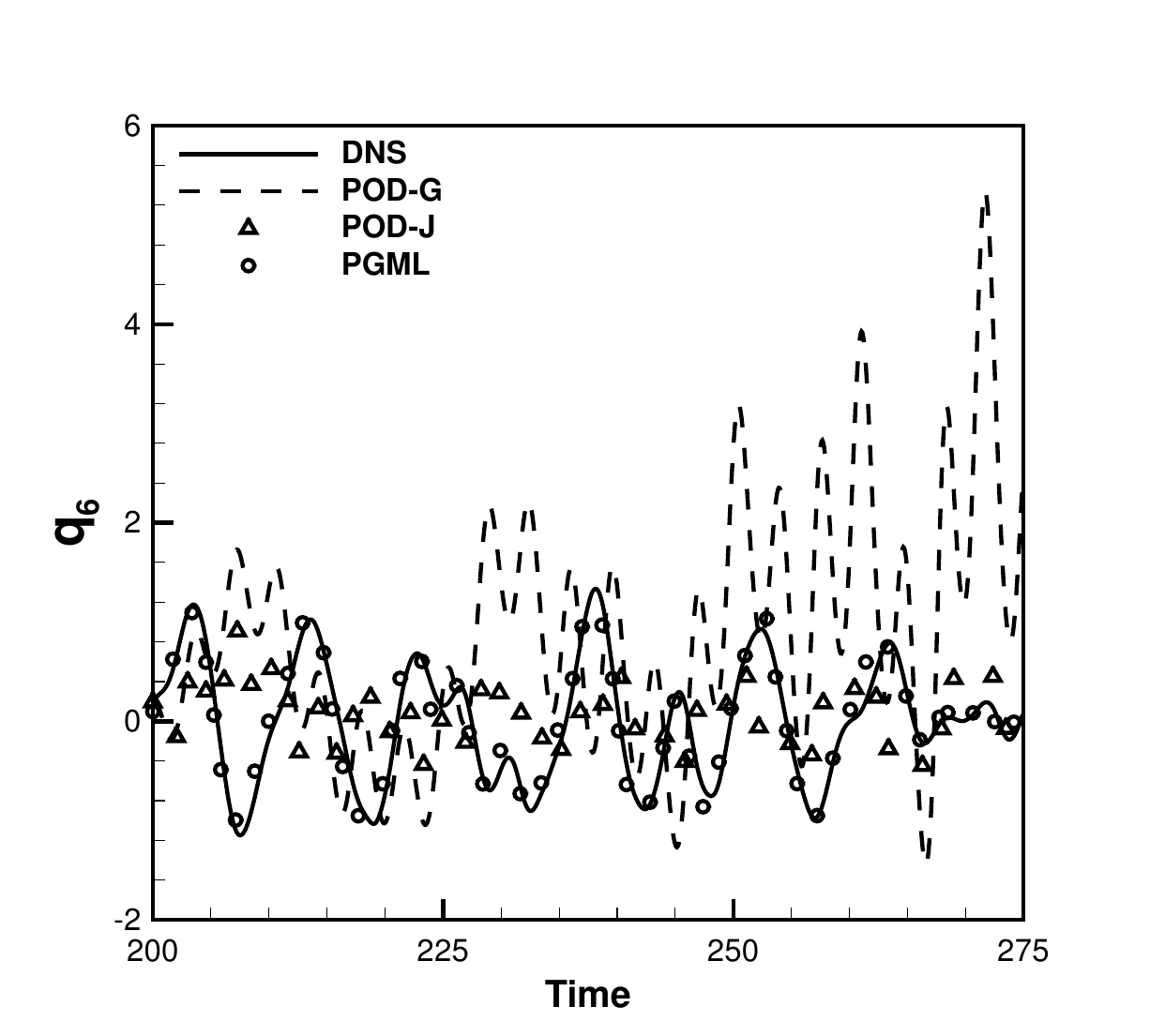}}
\caption{The prediction of temporal coefficients for six POD modes: DNS (solid line), POD-G (dash line), POD-J (triangle), and PGML (circle).}
\label{fig:Coefficients_closure}
\end{figure}

\section{Galerkin Projection Methods}
The velocity flow field can be written in terms of mean and the fluctuating parts as follows:
\begin{equation}
\mathop{u}({\bf{x}},t) \approx {{\mathop{u}}_M}({\bf{x}},t) \equiv \overline {\mathop{u}} ({\bf{x}}) + \sum\limits_{i = 1}^M {{q_i}\left( t \right)} {\Phi _i}\left( {\bf{x}} \right)
\end{equation}
where the $q_i$ are the temporal coefficients of velocity, ${\mathop{u}_M}$ is the modeled velocity field, and $\overline {\mathop{u}}$ is the mean velocity field. In Galerkin projection method, the conventional ROM can be developed by projecting the Navier-Stokes equations onto the POD modes ($\Phi_k$) as
\begin{equation}
\left(\mathop{u_t},\Phi_k \right) - \left(\nu \Delta \mathop{u},\Phi_k\right) + \left(\left( {\mathop{u}\cdot\nabla } \right)\mathop{u},\Phi_k\right) + \left(\nabla p,\Phi_k\right) = 0
\label{eq:POD-proj}
\end{equation}
In this study, we consider an external flow field and a large computational domain therefore, the pressure term does not appear in the ROM (for details, see \cite{noack2005need}). As a result, equation (\ref{eq:POD-proj}) yields the following nonlinear autonomous dynamical system:
\begin{equation}
\dot q_k(t)=\mathcal{C}_k+\sum_{m=1}^{M} \mathcal{D}_{km}  q_m(t)+\sum_{m=1}^{M} \sum_{n=1}^{M} \mathcal{E}_{kmn}  q_m(t) q_n(t),
\label{eq:PODG}
\end{equation}
where
\begin{align*}
\mathcal{C}_{k}&= \nu \left(\Delta \bar{\mathbf{u}},{\Phi_k}\right)-\left(\left( {\bar{\mathbf{u}}\cdot\nabla } \right)\bar{\mathbf{u}},\Phi_k\right),
\\ \mathcal{D}_{km}&=-\left(\bar{\mathbf{u}} \cdot \nabla{\Phi}_{m},{\Phi}_k\right)-\left({\Phi}_m \cdot \nabla \bar{\mathbf{u}},{\Phi}_k\right)+ \nu \left(\Delta {{\Phi}_m},{\Phi_k}\right),
\\ \mathcal{E}_{kmn}&=-\left({\Phi}_m \cdot \nabla{\Phi}_{n},{\Phi}_k\right).
\end{align*}
The conventional ROM is also termed as POD-G model and provides promising results for laminar flows. However, it may unstable for turbulent flows where the higher modes contain a relevant amount of energy, as shown in Fig. \ref{fig:eigen}. Therefore, the effect of higher-index modes is incorporated through the closure modeling for an effective Galerkin projection ROM in turbulent flows.

Nonlinear closure models in ROM are inspired by the existing CFD techniques in turbulence because the concept of eddy-viscosity can be translated into the ROM settings. Here, we will consider the Jacobian nonlinear closure model that is termed as POD-J model. This model can be written in a mathematical form as
\begin{align}
{\dot q_k}(t) &= \left({\mathcal{C}_k}+\mathcal{\tilde C}_{k}\right) + \sum\limits_{m = 1}^M {\left( {{\mathcal{D}_{km}} + {\mathcal{\tilde D}_{km}}} \right)} {q_m}(t) \nonumber \\
&+ \sum\limits_{m = 1}^M {\sum\limits_{n = 1}^M {{\mathcal{E}_{kmn}}} } {q_m}(t){q_n}(t).
\label{eq:PODJ}
\end{align}
where vector $\mathcal{\tilde C}_{k}$ and matrix ${\mathcal{\tilde D}_{km}}$ depend on the temporal coefficients $\bf{q}$ from previous time step that can be written as
\begin{align}
\mathcal{\tilde C}_{k}  &= \nu_J \left(\Delta \bar{\mathbf{u}},{\Phi_k}\right), \nonumber \\
\mathcal{\tilde D}_{km} &= \nu_J \left(\Delta {{\Phi}_m},{\Phi_k}\right). \nonumber
\end{align}

The eigenvalues of $\mathcal{\tilde D}_{km}$ are negative and, therefore, it acts as a stabilizing agent for the POD-G model. In other words, a dissipation term is added that increases the viscosity coefficient by $\nu_J = \left(C_s \delta\right)^2\left| {{S^J}}\right|$ similar to the Smagorinsky model in the LES approach. POD-J employs the dynamical system approach to compute the deformation tensor $\left| {{S^J}}\right|$. In this approach, we determine the Jacobian of the POD-G model by differentiating the POD-G with respect to its states as:
\begin{equation}
\frac{{\partial{{\dot q}_k}}}{{\partial{q_l}}} = {\mathcal{D}_{kl}} + \sum\limits_{m = 1}^M {{\mathcal{E}_{klm}}} {q_m} + \sum\limits_{n = 1}^M {{\mathcal{E}_{kln}}} {q_n}
\end{equation}
where $\left| {{S^J}}\right|$ is Frobenius matrix norm of the Jacobian of the POD-G model, i.e. Frobenius matrix norm of $\frac{{\partial{{\dot q}_k}}}{{\partial{q_l}}}$. The key advantage of POD-J model is that it can be evaluated with a negligible computational cost at each time step \cite{imtiaz2020nonlinear}.
\section{Physics-Guided Machine Learning Model}
In this section, we describe the various elements of proposed PGML framework for developing a reduced-order representation of turbulent flows.  We first present the general form of the machine learning model which is trained by the time series of observable  $\left\{ {\boldsymbol{o}^{(1)} ,\boldsymbol{o}^{(2)}, ...,\boldsymbol{o}^{(N)} } \right\}$ where $\boldsymbol{o}\in \mathbb{R}^{do}$ are sampled at regular intervals. The general form of the machine learning model can be written as

\begin{equation}
\boldsymbol{h}^{\left( t \right)}  = F_h^h \left(\boldsymbol{o}^{(t)} ,\boldsymbol{h}^{\left( {t - 1} \right)}  \right),
\end{equation}
\begin{equation}
\boldsymbol{{\tilde{o}}}^{ \;\left( t+1 \right)}  = F_h^o \left( {\boldsymbol{h}^{\left( {t} \right)} } \right),
\end{equation}
where $\boldsymbol{{\tilde{o}}}^{ \;\left( t+1 \right)}$ is forecast by using a machine learning model for the next observable state $\boldsymbol{{o}}^{ \left( t+1 \right)}$, $\boldsymbol{{o}}^{ \left( t \right)}$ is current observable state, $\boldsymbol{h}^{\left( {t} \right)}\in \mathbb{R}^{d_h }$ is the hidden state,  $F_h^h$ is the hidden-to-hidden mapping and $F_h^o$ is the hidden-to-output mapping.
The gating mechanism in LSTM cell is employed to mitigate the issue with vanishing (or exploding) gradient. This gating mechanism allows the information to be forgotten for time series prediction. The equations for forecast hidden state in terms of gating functions and input vectors can be written as
\begin{equation}
h^{\left( t \right)}  = G_o^t  \odot \tanh \left( {C_e^{\left( t \right)} } \right),
\end{equation}
where
\begin{equation*}
C_e^{\left( t \right)}  = G_f^t  \odot C_e^{\left( {t - 1} \right)}  + G_i^t  \odot \widetilde{C_e}^{\left( t \right)},
\end{equation*}
\begin{equation*}
\widetilde{C_e}^{\left( t \right)}  = \tanh \left[ {w_c \left[ {\boldsymbol{h}^{\left( {t - 1} \right)} ,\boldsymbol{o}^{\left( t \right)} } \right] + B_c } \right],
\end{equation*}
\begin{equation*}
G_f^t  = \sigma _f \left[ {w_f \left[ {\boldsymbol{h}^{\left( {t - 1} \right)} ,\boldsymbol{o}^{\left( t \right)} } \right] + B_f } \right],
\end{equation*}
\begin{equation*}
G_i^t  = \sigma _i \left[ {w_i \left[ {\boldsymbol{h}^{\left( {t - 1} \right)} ,\boldsymbol{o}^{\left( t \right)} } \right] + B_i } \right],
\end{equation*}
\begin{equation*}
G_o^t  = \sigma _o \left[ {w_o \left[ {\boldsymbol{h}^{\left( {t - 1} \right)} ,\boldsymbol{o}^{\left( t \right)} } \right] + B_o } \right].
\end{equation*}
Here $G_i^t$, $G_o^t$, $G_f^t$ $\in \mathbb{R}^{d_h }$ are input, output, and forget gates, respectively. $C_e^{\left( t \right)}$ is the cell state, whereas $w_f$, $w_c$, $w_i$, $w_o\in \mathbb{R}^{d_h \times (d_h+d_0)}$ are weight matrices and $B_f$, $B_c$, $B_i$, $B_o\in \mathbb{R}^{d_h }$ are bias vectors. $\sigma _i$, $\sigma _o$, $\sigma _f$ are sigmoid activation functions and $\odot$ represents the element wise multiplication. We can write the hidden-to-output mapping as
\begin{equation}
\boldsymbol{{\tilde{o}}}^{ \;\left( t+1 \right)}  = w_o{\boldsymbol{h}^{\left( {t} \right)} }
\end{equation}
where $w_o\in \mathbb{R}^{do \times d_h}$ is a linear activation function. Figure \ref{fig:Comparison} presents the schematic of the PGML model where the output of the POD-J model is concatenated into the middle layers of neural networks without breaking the neural network training process. In other words, the injection of known physics can be done in PGML which reduces the uncertainty of the ROM. LSTM cells are employed for hidden layers as they possess capability to model the long-term temporal dependencies without suffering from the vanishing gradient problem of recurrent neural network \cite{hochreiter1997long}.

In the PGML model, we first employ $(w-1)$ LSTM layers in PGML model and embedded the extracted features from POD-J into  $w$ intermediate layer as
\begin{align}
{\cal F}_{PGML} \left( {\xi ;\theta } \right) &= {\bf{h}}_{N_l } \left( {.;\Theta _{N_l } } \right) \circ \ldots \circ C\left( {{\bf{h}}_w \left( {.;\Theta _w } \right),\bf{q}^{\left( t \right):\left( {t - d + 1} \right)} } \right) \nonumber \\
&\quad \circ \ldots \circ {\bf{h}}_2 \left( {.;\Theta _2 } \right) \circ {\bf{h}}_1 \left( {\xi,\Theta _1 } \right)
\end{align}
The output of each LSTM layer ($i=1,\; 2,....,\; w-1,\; w+1,....\;N_l-1$) is $  {\bf{h}}_i \left( {.;\Theta _i }\right)\in \mathbb{R}^{d\times N_h }$ where $N_h$ is the number of hidden units in the LSTM cells. The last layer maps the final hidden state to output:  $ {\bf{h}}_{N_l } \left( {.;\Theta _{N_l } } \right): \mathbb{R}^{N_h }\xrightarrow{}\mathbb{R}^{M}$.  $ C\left(. , .\right) $ represents the concatenation operation between $w$th hidden layer and the temporal coefficients from POD-J model . Therefore, the output of $w\text{th}$ intermediate layer will be $\mathbb{R}^{d \times (M+N_h) }$ as the output of POD-J is embedded only in the single layer of the neural network. It is important to note that POD-J assists the LSTM network in constraining the output to a manifold of the physically realizable solution that will improve its generalization for extrapolation regime.

\section{Numerical Results and Discussions}
We determine the essential features of the POD-G, POD-J, and PGML models by developing the ROM for flow past a circular cylinder at $\rm{Re}=1000$ with a turbulent wake. During the post-processing, we place the rectangular domain ($L=9.5$, $W=24$, $H=2$) in front of a circular cylinder and interpolate the velocity field on it. We consider the following criteria for comparing the performance of various ROMs: the temporal coefficients ($q_i$), the mean value of the velocity field, Reynolds stresses, and the kinetic energy spectrum.
\begin{figure}[htbp]
\centering
\subfigure[]{}
{\includegraphics[angle=0,width=0.49\linewidth]{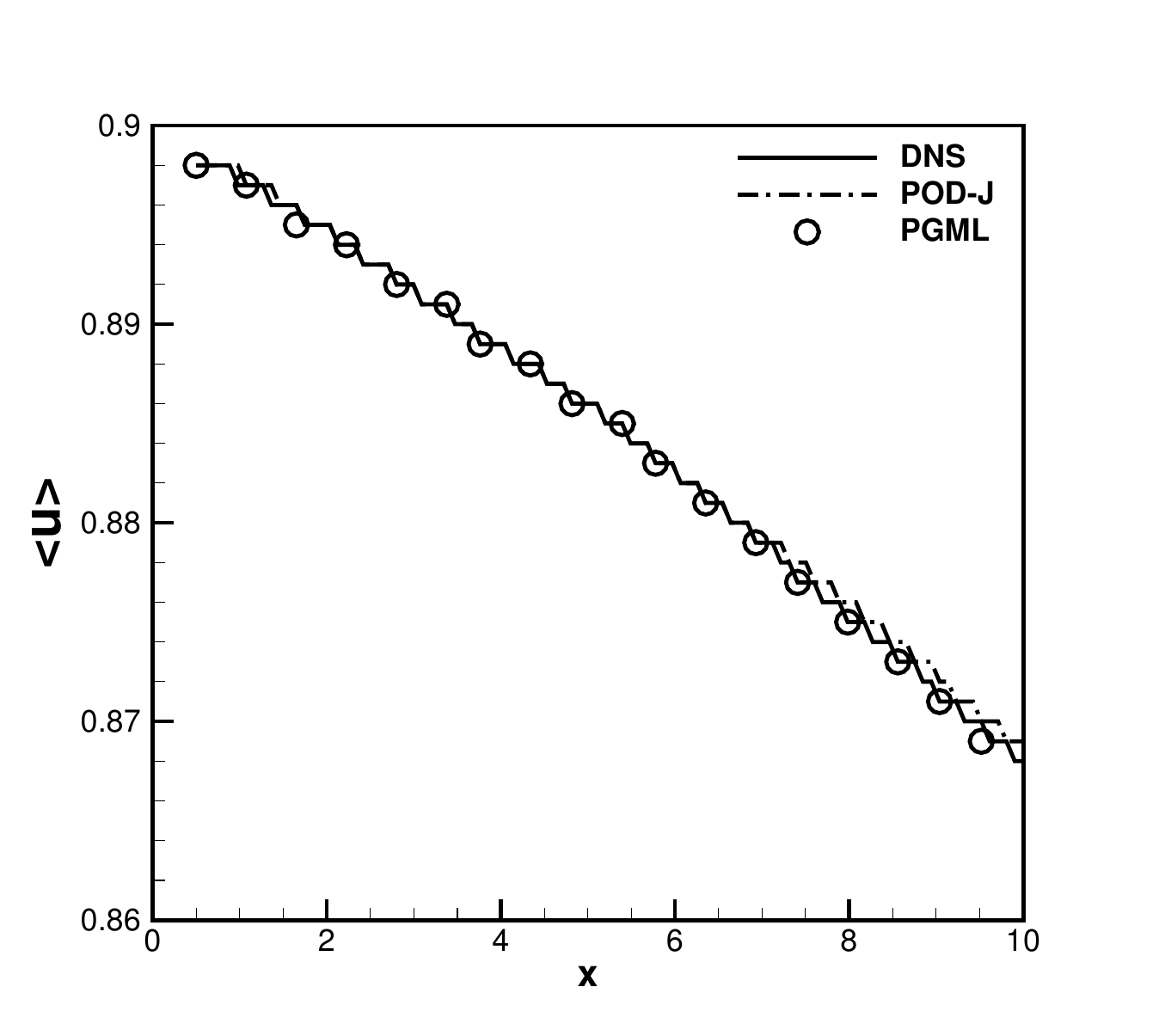}}
\subfigure[]{}
{\includegraphics[angle=0,width=0.49\linewidth]{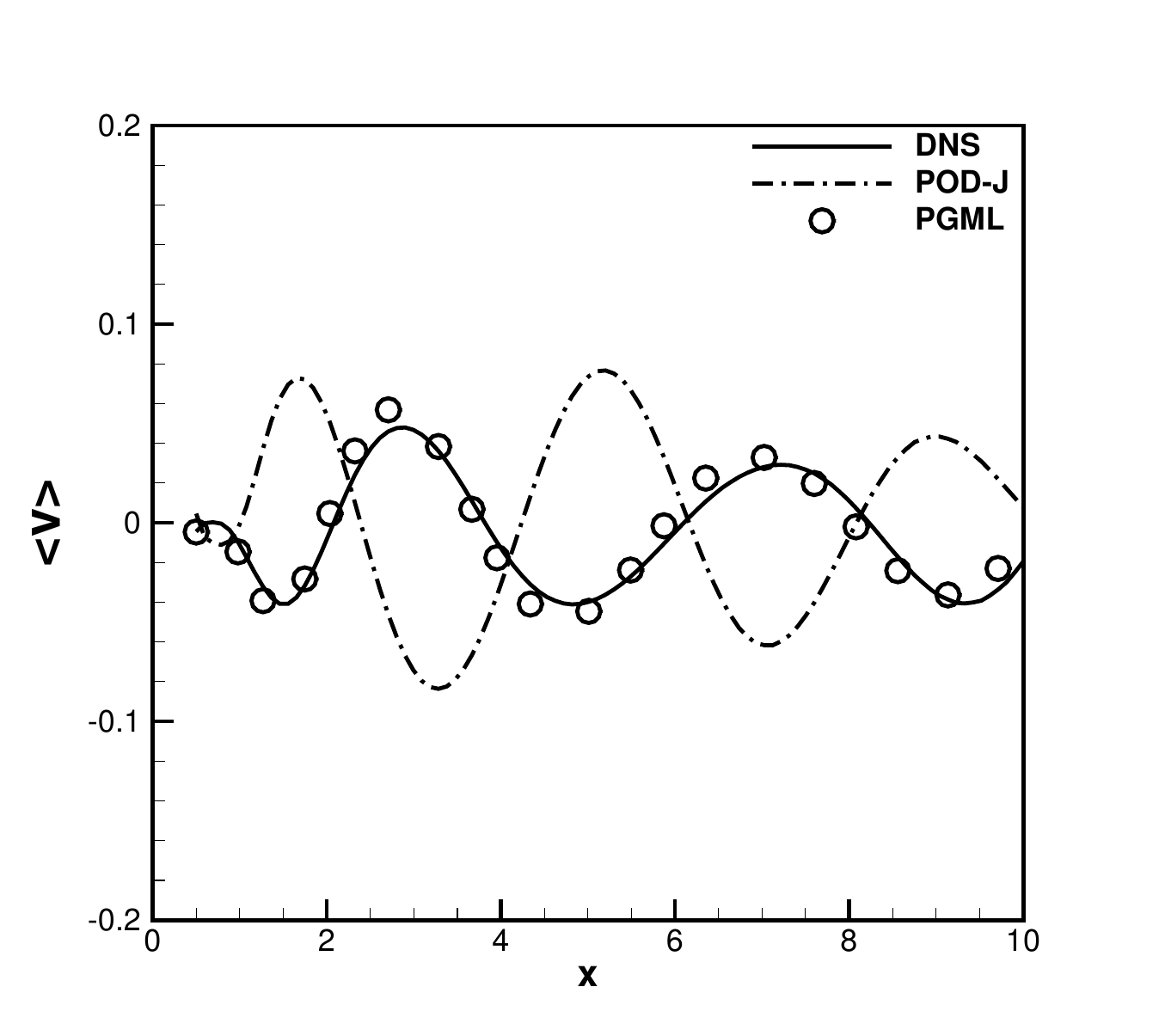}}
\subfigure[]{}
{\includegraphics[angle=0,width=0.49\linewidth]{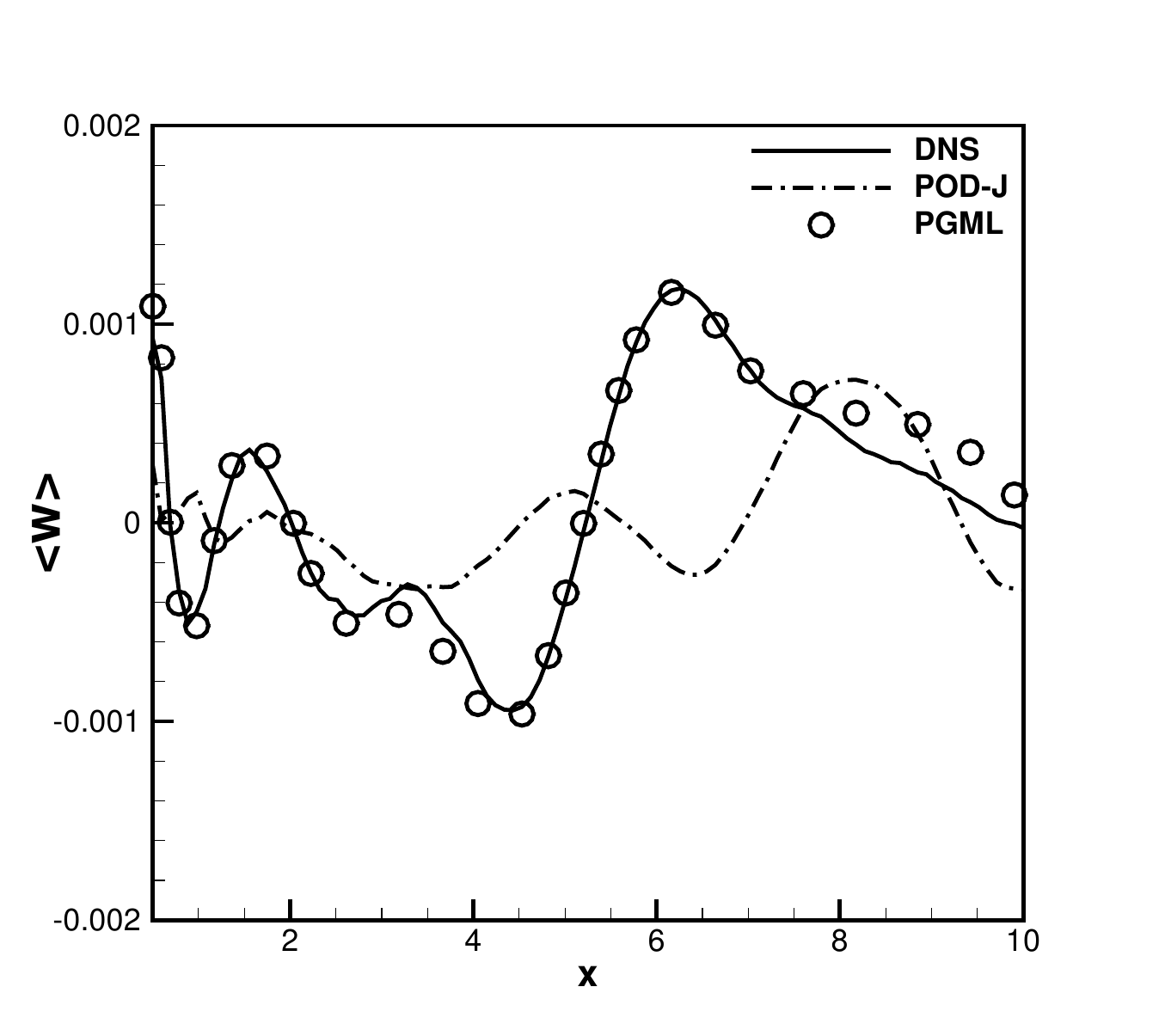}}
\caption{ Mean value of velocity field for DNS (solid line), POD-J (dash dotted line), and PGML (circle).}
\label{fig:Mean_velocity}
\end{figure}
\FloatBarrier
\begin{figure}[htbp]
\centering
\subfigure[POD-J]
{\includegraphics[angle=0,width=0.45\linewidth]{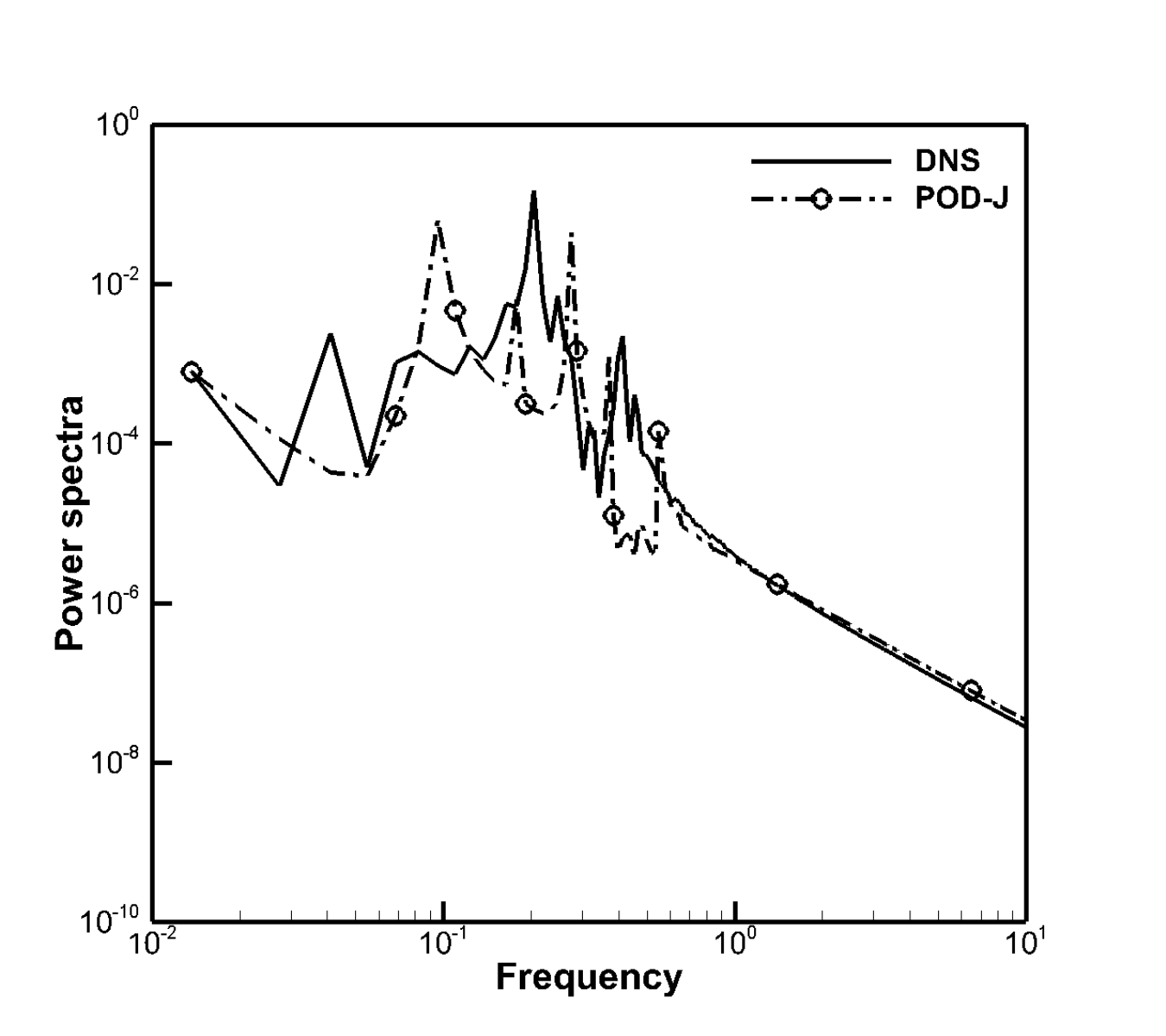}}
\subfigure[PGML]
{\includegraphics[angle=0,width=0.45\linewidth]{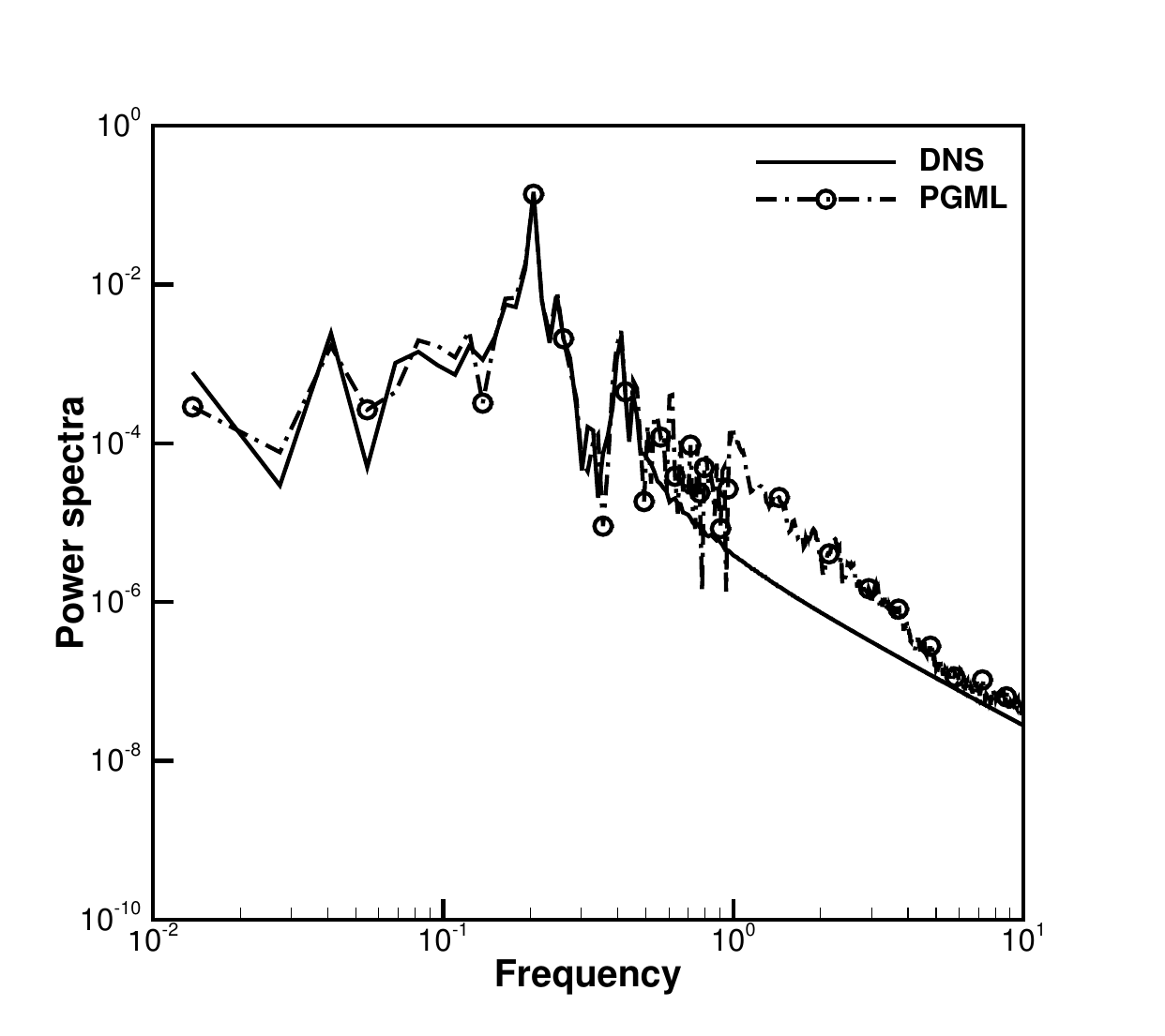}}
\caption{Kinetic energy spectrum: a) POD-J and b) PGML models.}
\label{fig:FFT}
\end{figure}
\subsection{Temporal Coefficients}
PGML network consists of six layers with the first five layers as the LSTM layers and the last dense layer as the output layer. Here the temporal coefficients from POD-J model are concatenated in the third layer.  It is important to note that the optimal layer for embedding these physics-based features is another critical hyperparameter in the PGML framework. Each LSTM layer is configured with 30 hidden units and employs the hyperbolic tangent (\textit{tanh}) activation function for cell state updates and the sigmoid activation functions for the gating mechanisms. We find through numerical experimentation that the deeper network architecture does not improve the results further. On the other hand, it may sometimes lead to poor performance due to over-fitting.

For intrusive ROM, we perform the Galerkin projection of the governing onto POD modes and determine  $\mathcal{C}_k$ , ${\mathcal{D}_{km}}$, and ${{\mathcal{E}_{kmn}}}$, where $k$, $m$, $n$ = 1, 2,..., 6. We then compute $q_i$ for POD-G model by using equation (\ref{eq:PODG}) and the ode45 function in MATLAB. Similar procedure is done with equation (\ref{eq:PODJ}) to compute $q_i$ for POD-J model. It is important to note that $\mathcal{C}_k$, ${\mathcal{D}_{km}}$, and ${{\mathcal{E}_{kmn}}}$ are computed a priori whereas $\mathcal{\tilde C}_{k}$ and $\mathcal{\tilde D}_{km}$ are computed at each time step of integration.

For PGML model, we consider three previous states of the system and employ them to predict the future state. During online deployment, we start with initial conditions for the first $d$ time steps. This information is used to predict the forecast state at the $(d+1)^{th}$ time step. Subsequently, the system's state at $(d+1)^{th}$ is then used to predict the forecast state at the $(d+2)^{th}$ time step. This process continues sequentially until the final time step is reached. Figure \ref{fig:Coefficients_closure} shows the time evolution of the temporal coefficients using the POD-G, the POD-J, and the PGML models. It also presents \textit{true} temporal coefficients, i.e. DNS, for comparison purposes, which are obtained by projecting the snapshot data onto the POD modes. The temporal coefficients from the POD-G model show a larger divergence from \textit{true} coefficients. The POD-J model improves the prediction accuracy of temporal coefficients but its prediction is not reliable, especially for higher-index POD mode. On the other hand, the prediction accuracy of the PGML model is better than the POD-J model even for higher temporal coefficients. Overall, the prediction accuracy of the POD-G model is not good as compared to the POD-J and PGML models. Therefore, the rest of the paper will focus exclusively on the POD-J and PGML models and compare their performance using various criteria.

\subsection{Mean Velocity Field and Energy Spectra}
In this section, we determine $<u>$ (where $<.>=<.>_{tyz}$), $<v>$, and $<w>$ on the rectangular domain for POD-J and PGML models. ROMs results are compared with mean velocity components from DNS data which are produced by the Galerkin projection of true temporal coefficients and the POD modes. It can be seen from Fig. \ref{fig:Mean_velocity} that $<u>$ can be predicted with reasonable accuracy for POD-J and PGML models. However, the prediction accuracy of PGML model is better than the POD-J model for normal and spanwise velocity components.

For further analysis, we compute the kinetic energy spectrum for both ROMs and compared their results with the DNS data as shown in Fig. \ref{fig:FFT}. Here, we place a probe within flow field (1.08, 0.51, 1.06) to monitor the intensity of its kinetic energy. The figure shows a satisfactory agreement between PGML and DNS results. On the other hand, the prediction accuracy of POD-J model is low because some additional peaks are observed as compared to DNS data. Furthermore, PGML model accurately captures the energy cascade region, which is a crucial characteristic of turbulent flows.
\subsection{Reynolds Stresses}
\begin{figure}[htbp]
\centering
\subfigure[]{\includegraphics[angle=0,width=0.45\linewidth]{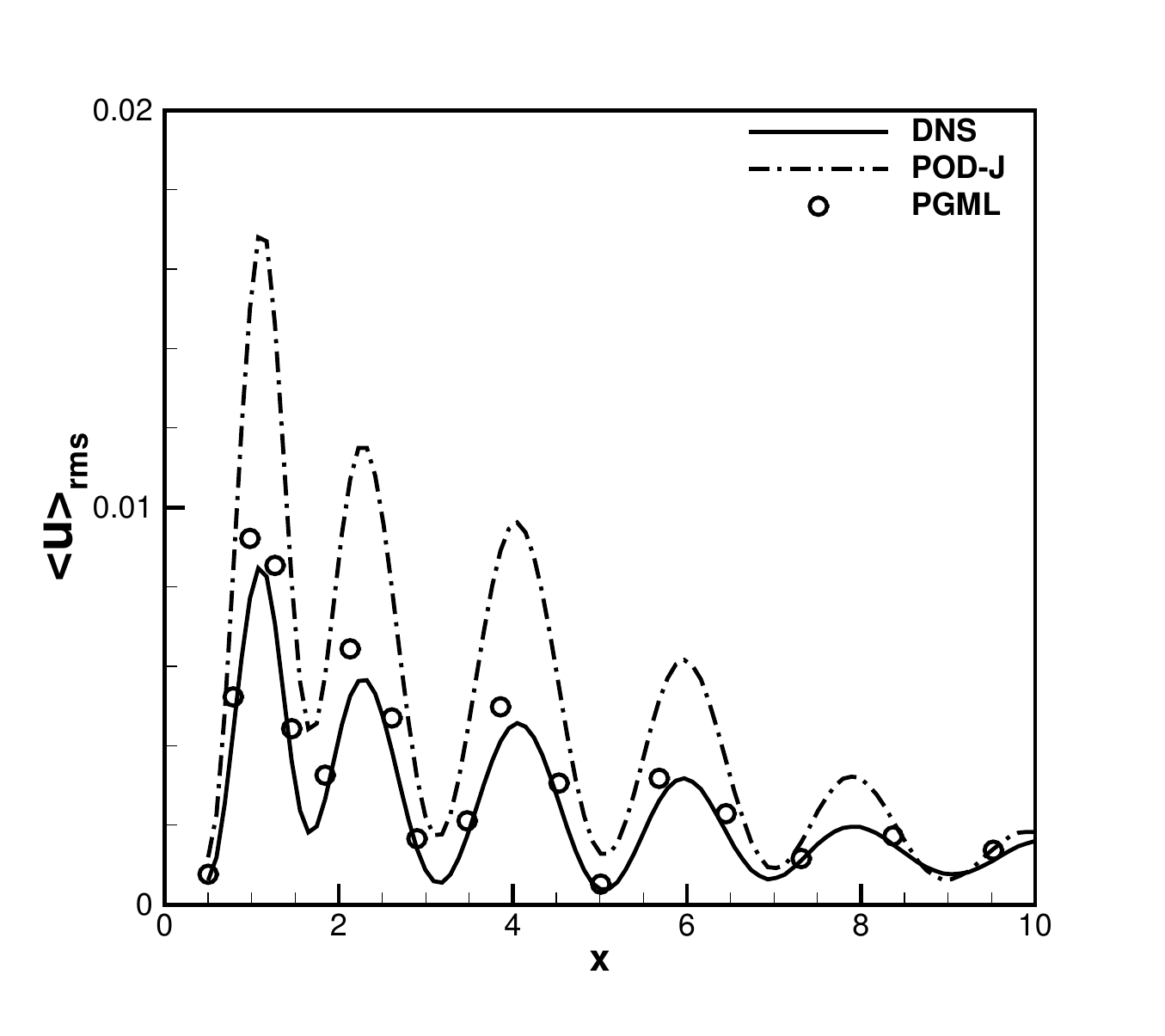}}
\subfigure[]{\includegraphics[angle=0,width=0.45\linewidth]{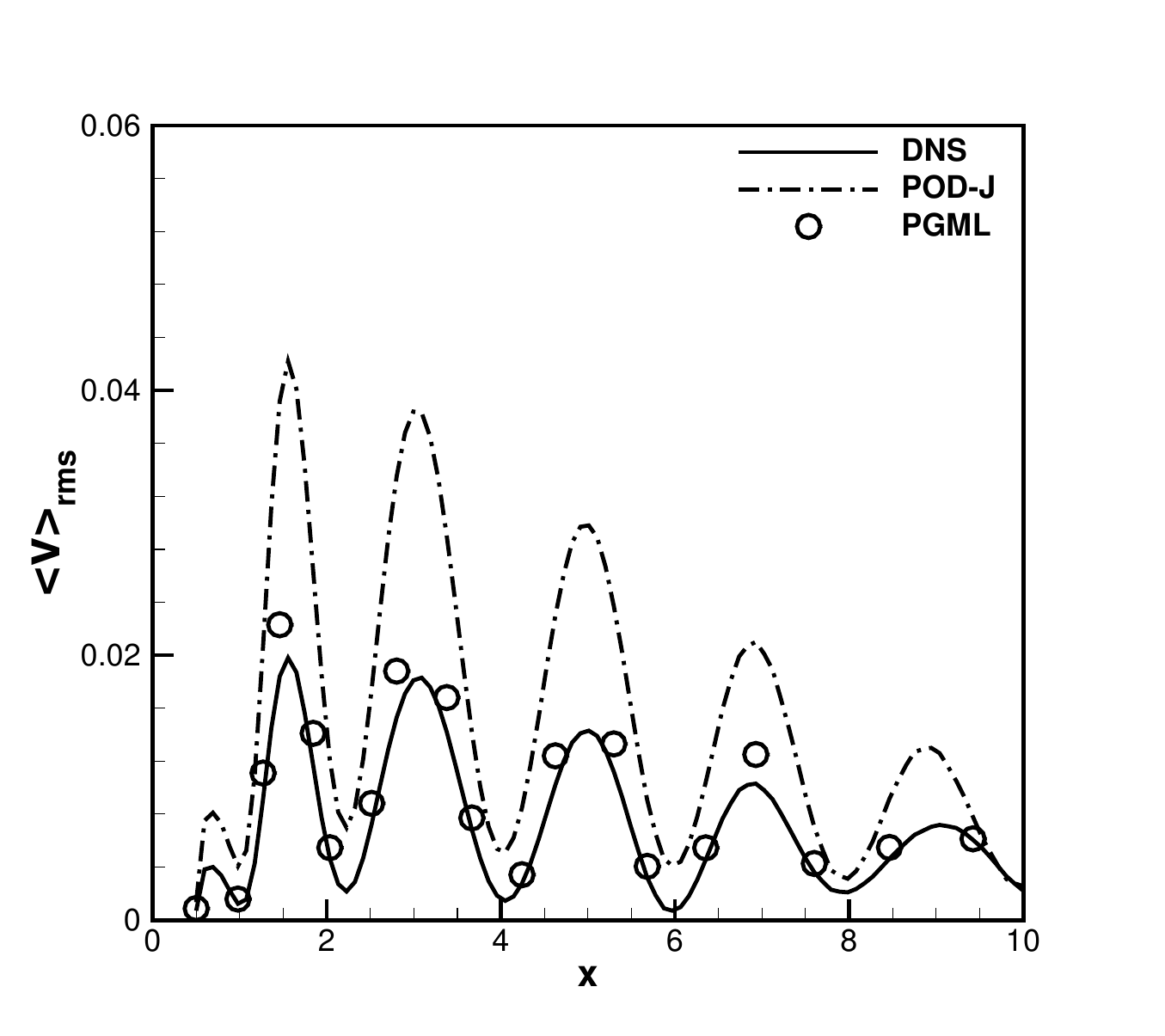}}
\subfigure[]{\includegraphics[angle=0,width=0.45\linewidth]{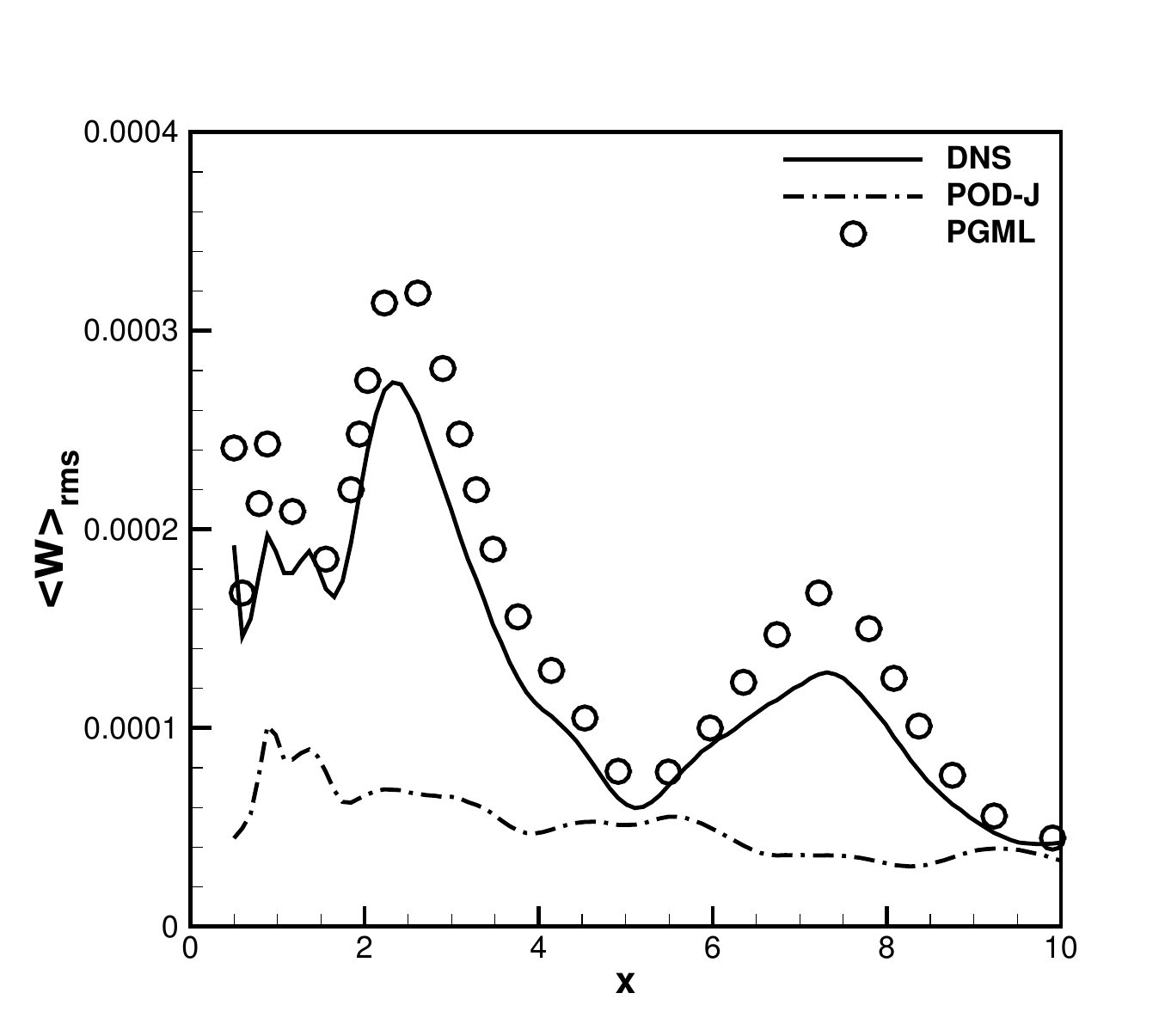}}
\caption{ Rms value of fluctuation in velocity components for DNS (solid line), POD-J (dash dotted line), and PGML (circle).}
\label{fig:RMS_velocity}
\end{figure}

The Reynolds stresses represent the components of total stress tensor in a fluid, accounting for turbulent fluctuations in fluid momentum. These stresses are symmetric, therefore we only determine the six components instead of nine components. The first three components of Reynolds stresses are the rms of the fluctuation in velocity components: $<u>_{rms}=<u-<u>,u-<u>>$; $<v>_{rms}=<v-<v>,v-<v>>$; and $<w>_{rms}=<w-<w>,w-<w>>$. Figure \ref{fig:RMS_velocity} shows the rms of the fluctuation in velocity components by using POD-J and the PGML models. The figure depicts that POD-J yields inaccurate results for rms values of all velocity components. Similar to the mean velocity case, the rms values of PGML model are accurate and closer to the rms obtained from DNS.
\FloatBarrier
\begin{figure}[htbp]
\centering
\subfigure[]{\includegraphics[angle=0,width=0.45\linewidth]{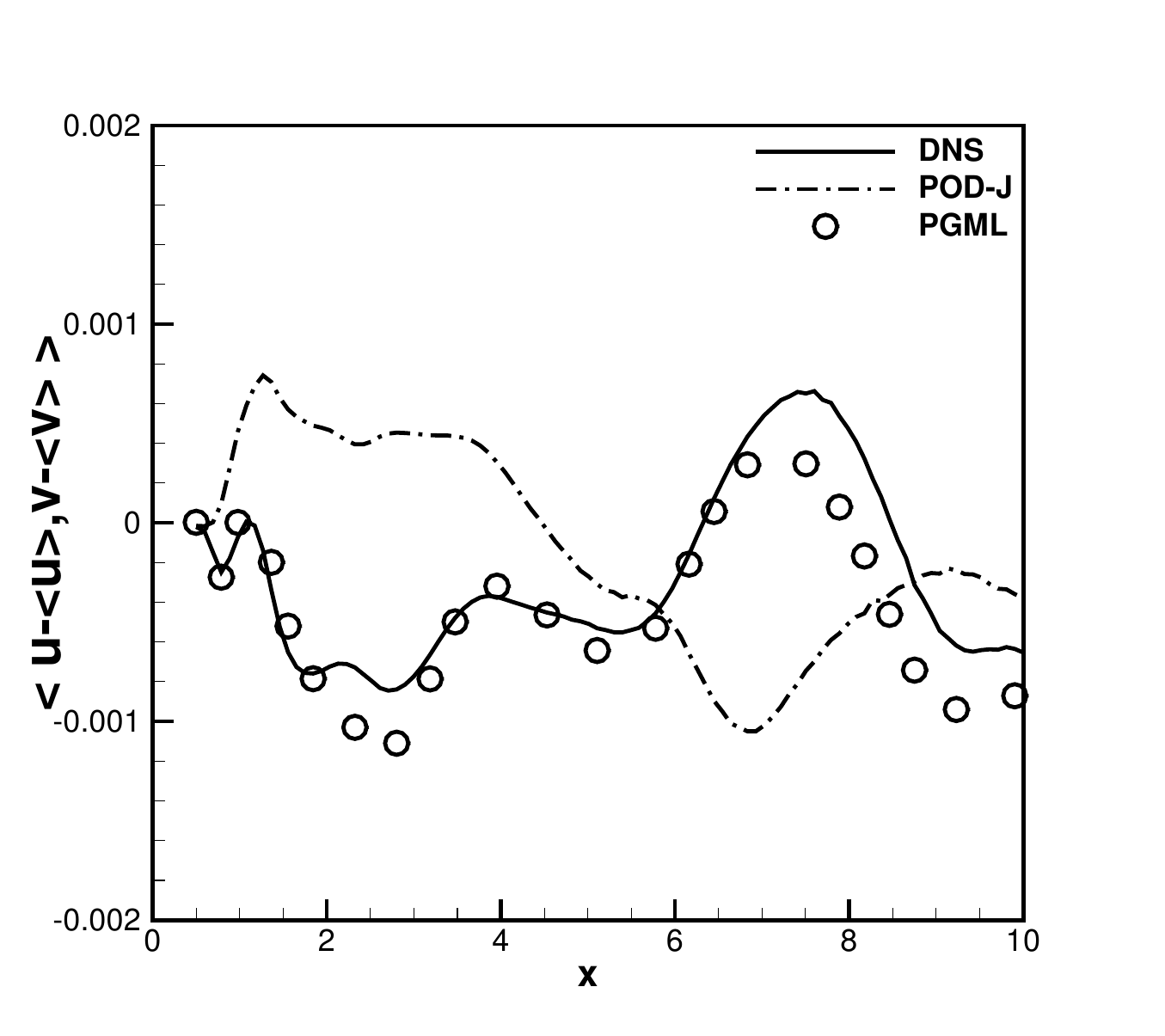}}
\subfigure[]{\includegraphics[angle=0,width=0.45\linewidth]{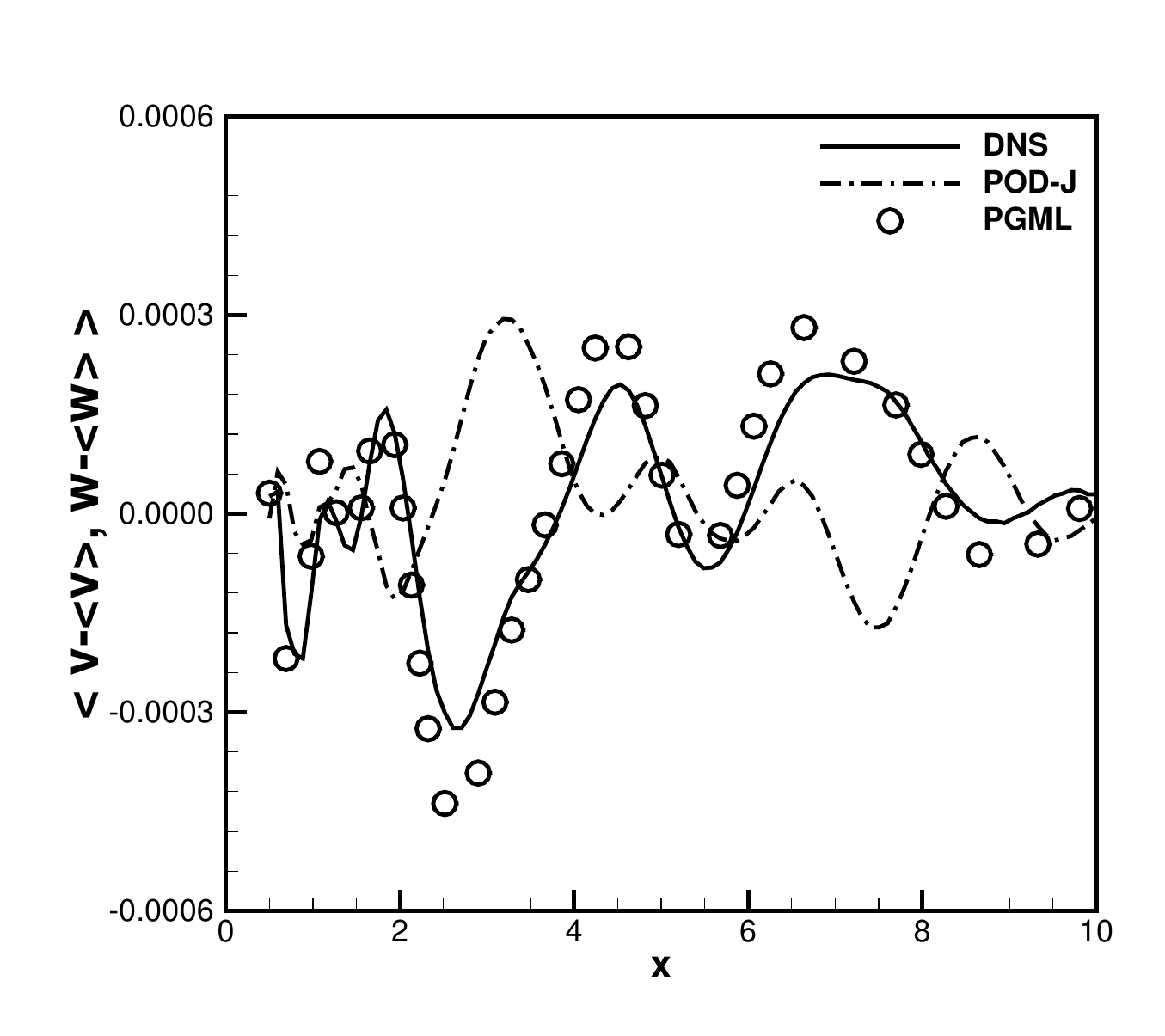}}
\subfigure[]{\includegraphics[angle=0,width=0.45\linewidth]{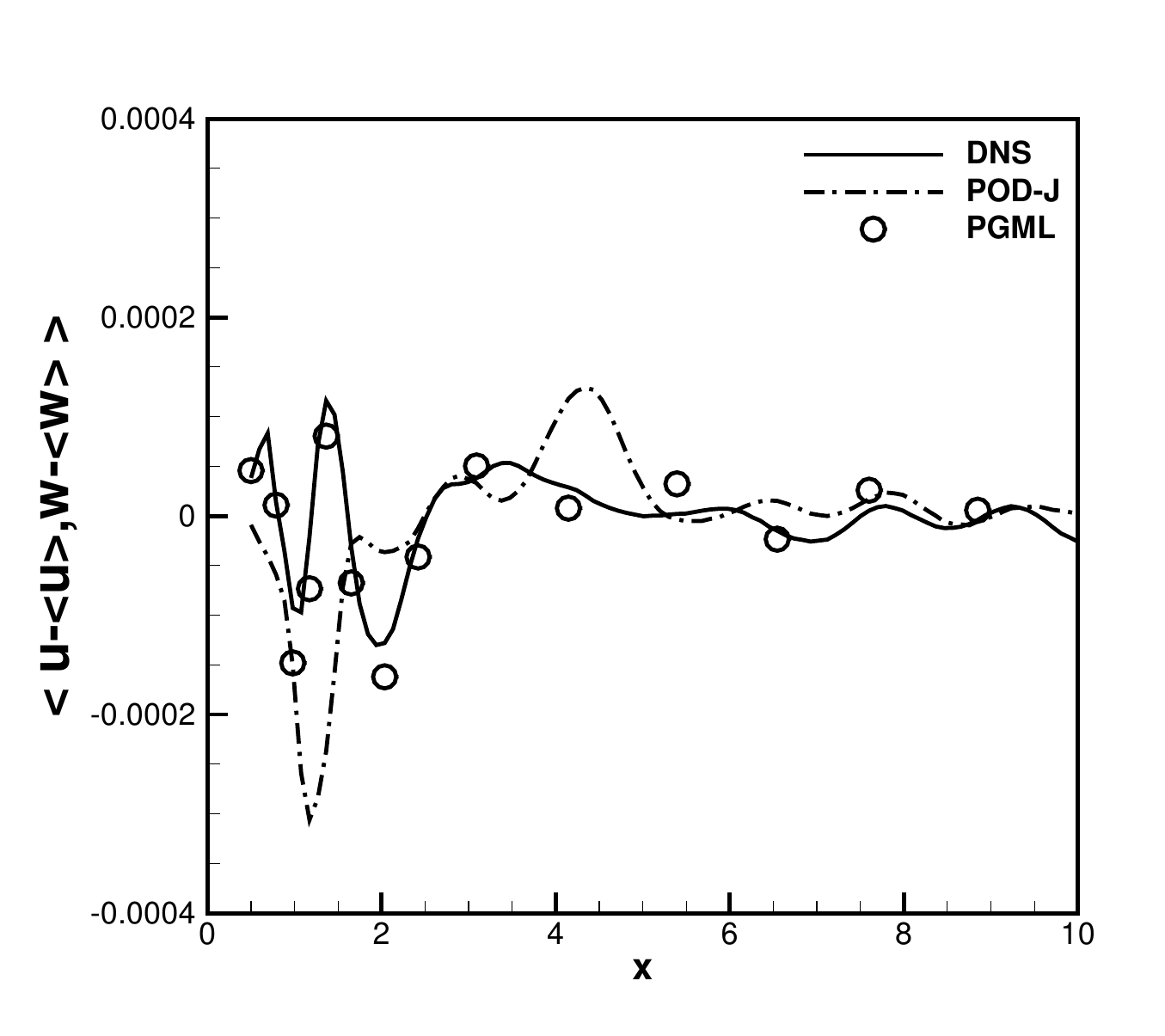}}
\caption{Reynolds stresses for DNS (solid line), POD-J (dash dotted line), and PGML (circle).}
\label{fig:Reynolds stresses}
\end{figure}

The remaining  symmetric Reynolds stress components are: $<u-<u>, v-<v>>$; $<v-<v>, w-<w>>$; and $<u-<u>, w-<w>>$. Figure \ref{fig:Reynolds stresses} depicts the Reynolds stresses for various values of $\rm{x}$. It reveals that the Reynolds stresses from the POD-J model consistently exhibit inaccuracies. In contrast, PGML enhances the distribution of Reynolds stresses, aligning more closely with the DNS data.

\begin{figure}[htbp]
\centering
\subfigure[]
{\includegraphics[angle=0,width=0.49\linewidth]{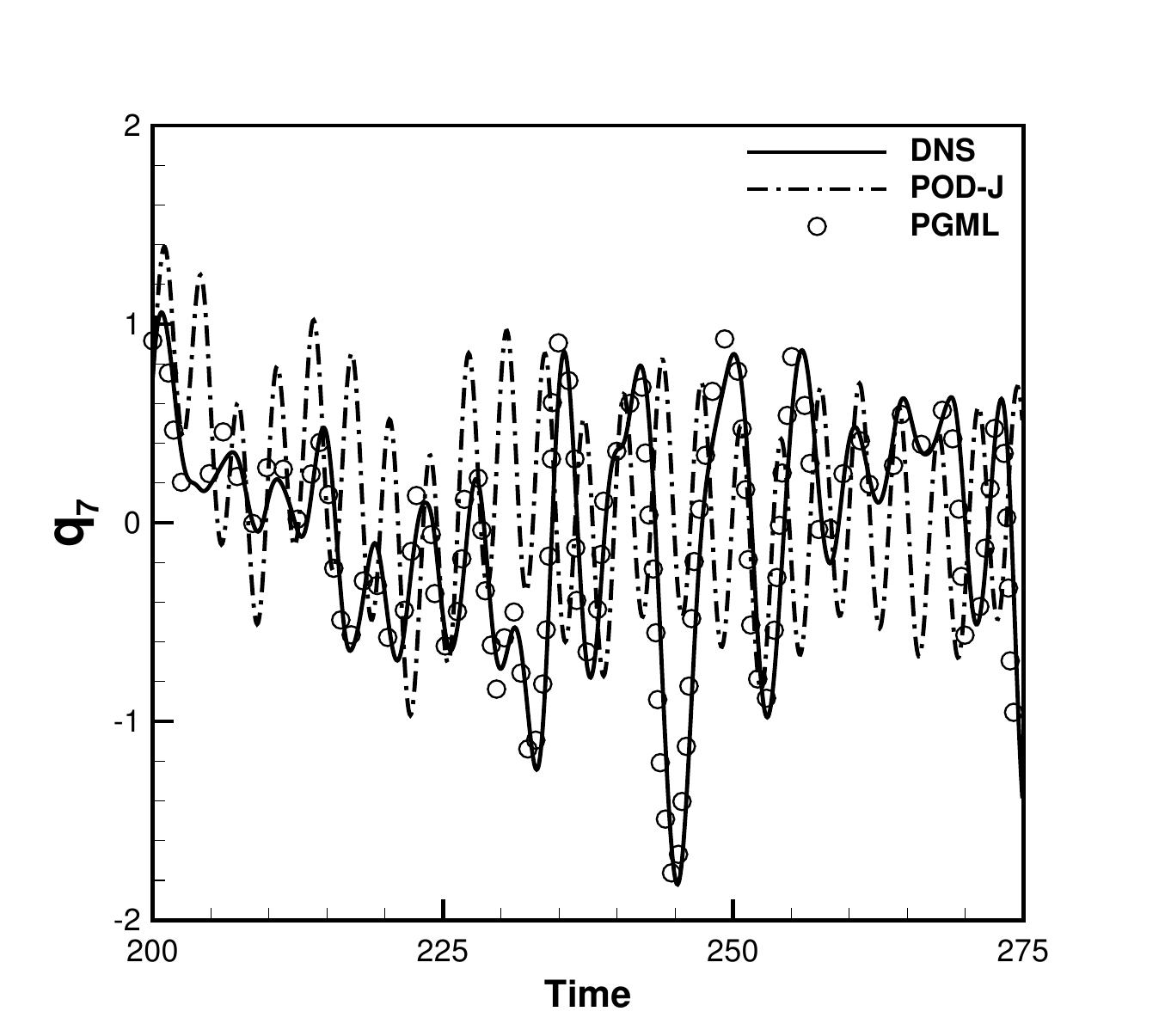}}
\subfigure[]
{\includegraphics[angle=0,width=0.49\linewidth]{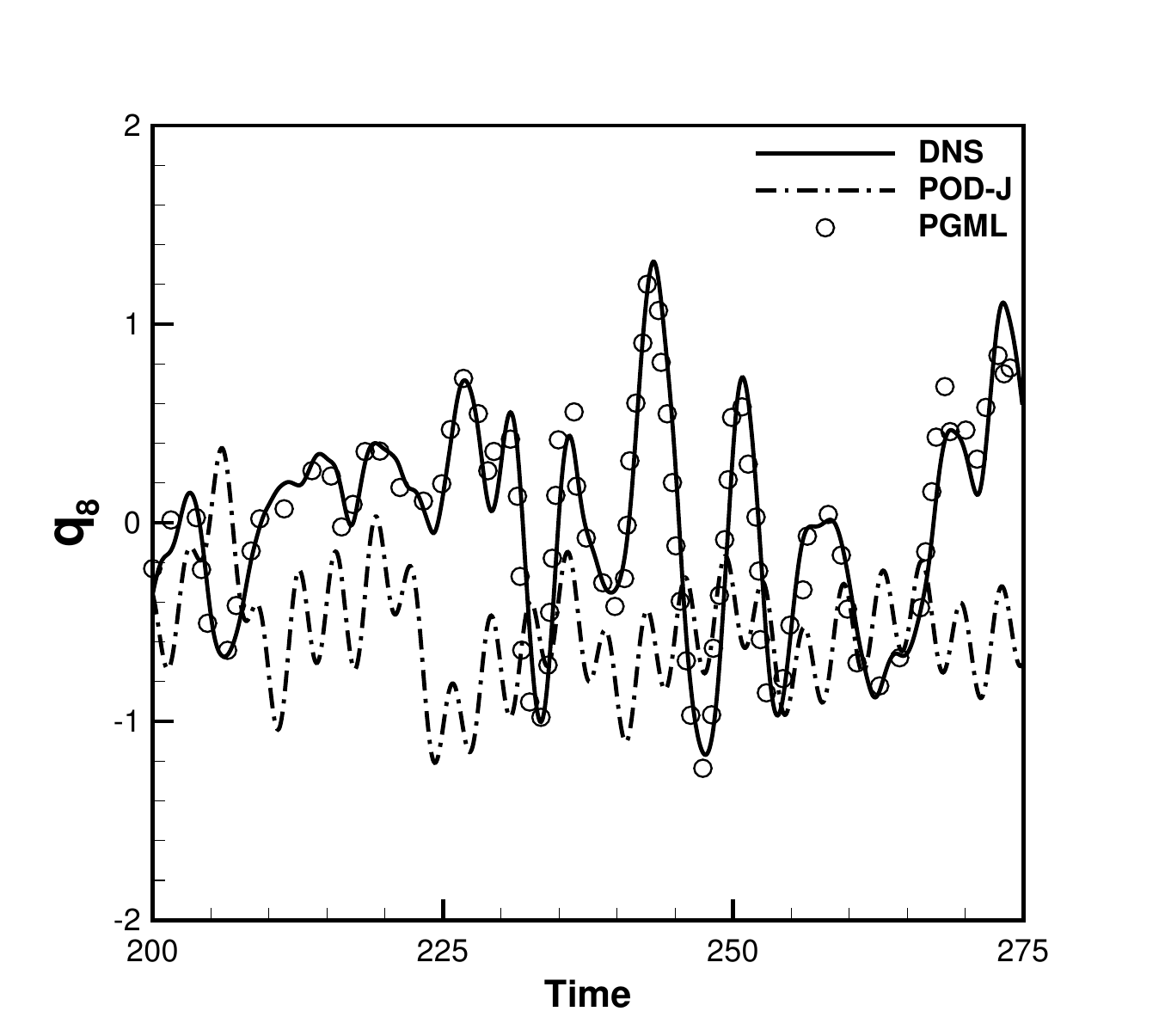}}
\subfigure[]
{\includegraphics[angle=0,width=0.49\linewidth]{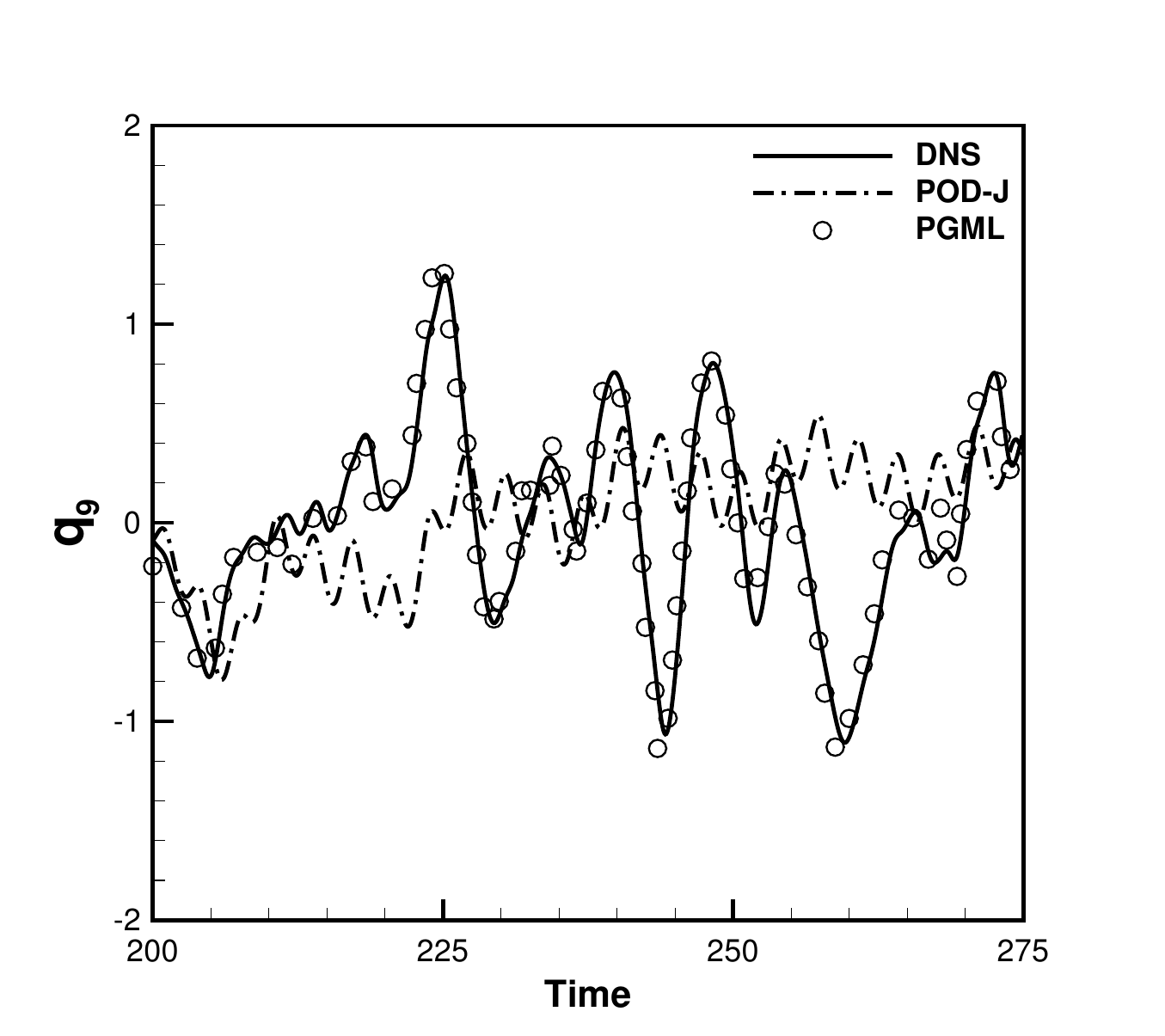}}
\subfigure[]
{\includegraphics[angle=0,width=0.49\linewidth]{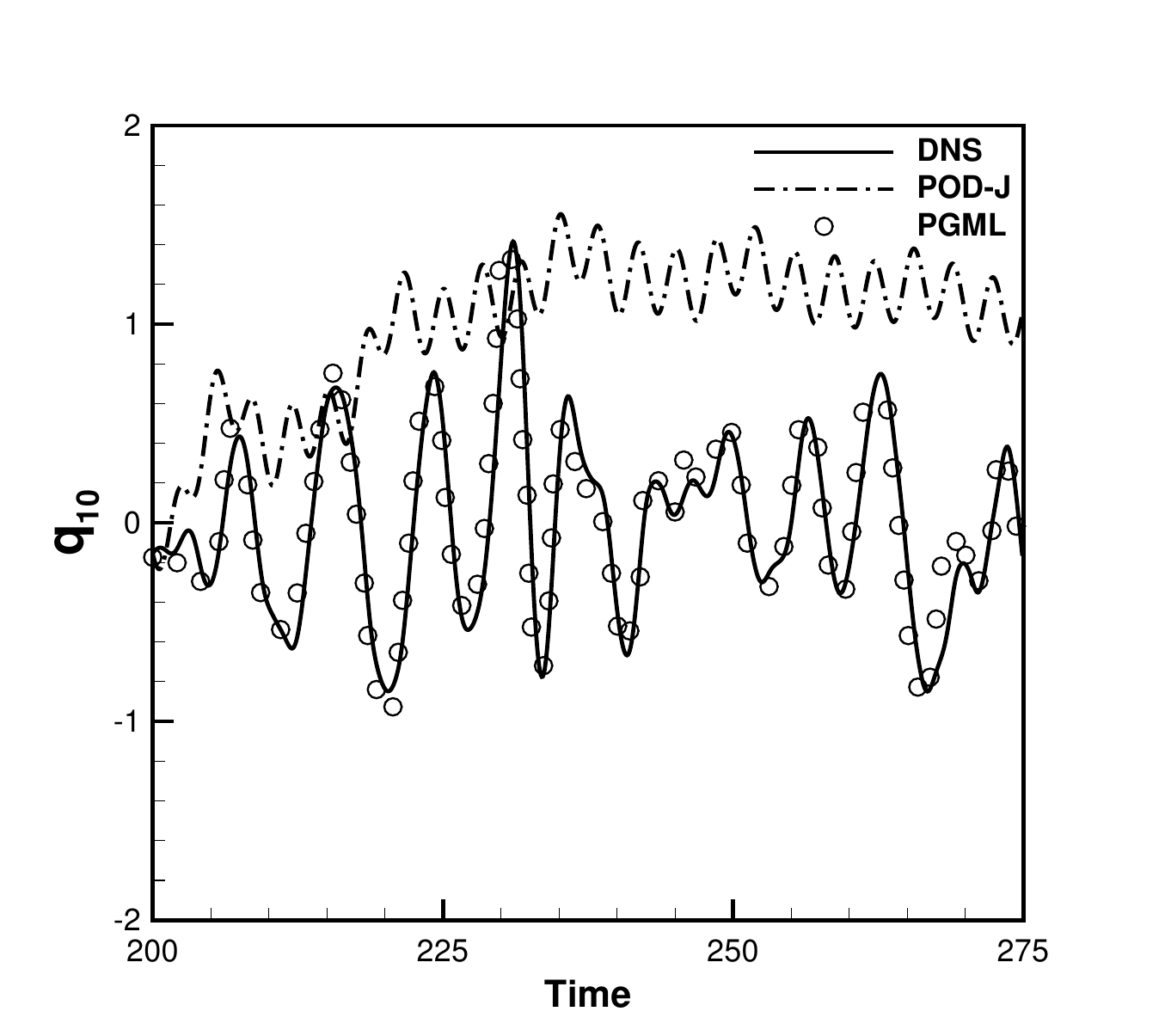}}
\caption{The prediction of temporal coefficients ($q_{7}$-$q_{10}$) for ten modes by using DNS (solid line), POD-J (dash-dot line), and PGML (circle).}
\label{fig:Coefficients_high}
\end{figure}
\subsection{Performance of ROM at Higher Modes}
The computational efficiency of the nonlinear closure model by using a dynamical system approach is reasonably good as compared to existing nonlinear closure models \cite{imtiaz2020nonlinear}. However, it is challenging to achieve better accuracy for coefficients corresponding to higher-index POD modes. In this section, we consider 10 POD modes and compared the nonlinear Jacobian ROM with the PGML model. Figure \ref{fig:Coefficients_high} shows the time evolution of the temporal coefficients ($i$ = 7, 8, 9, 10) using the POD-J and the PGML models. The temporal coefficients from the PGML model align more closely with the DNS data compared to those from the POD-J model. This indicates that the PGML model excels in predicting higher-index temporal coefficients, making it particularly well-suited for precise control applications.
\section{Conclusion}
The Galerkin ROM provides reasonable accuracy for lower-index temporal coefficients, but its prediction is not reliable for higher-index coefficients. In this study, we introduced a novel framework for a reliable prediction of higher-index coefficients in turbulent flows. We proposed the framework of the PGML model for turbulent flows by using the LSTM cells and the output from Galerkin projection-based nonlinear closure models. We tested the prediction accuracy of PGML model by simulating the flow past a cylinder with a turbulent wake. The result shows that the accuracy of the PGML model is higher as compared to Galerkin projection-based nonlinear closure models. Furthermore, the PGML model is reliable for the accurate prediction of higher-index temporal coefficients therefore, it can be effectively employed in industrial applications, where dimensional reduction is helpful or required as in control or design over a large space of parameters.

\newpage
\clearpage
\bibliographystyle{spmpsci}
\bibliography{main.bbl}
\end{document}